\documentclass[aps,prfluids,groupedaddress,amsmath,amssymb]{revtex4-2}
\usepackage{empheq}
\usepackage[percent]{overpic}
\usepackage{graphicx}% Include Fig. files
\usepackage{dcolumn}% Align table columns on decimal point
\usepackage{bm}% bold math
\usepackage{hyperref}% add hypertext capabilities
\usepackage[mathlines]{lineno}% Enable numbering of text and display math
\begin{document}

\preprint{APS/123-QED}

\title{Fingering instability of thin films with an inclined contact line}% Force line breaks with \\

\author{Ze-Hui Gao}
 % \altaffiliation{Physics Department, XYZ University.}%Lines break automatically or can be forced with \\
\author{Er-Qiang Li}
\author{Peng Gao}%
 \email[Contact author: ]{gaopeng@ustc.edu.cn}
\affiliation{State Key Laboratory of High Temperature Gas Dynamics, School of Engineering Science, University of Science and Technology of China, Hefei, Anhui 230026, PR China
}%

% \collaboration{MUSO Collaboration}%\noaffiliation

%  \homepage{http://www.Second.institution.edu/~Charlie.Author}
% \affiliation{
%  Second institution and/or address\\
%  This line break forced% with \\
% }%
% \affiliation{
%  Third institution, the second for Charlie Author
% \affiliation{%
%  Authors' institution and/or address\\
%  This line break forced with \textbackslash\textbackslash
% }%

% \collaboration{CLEO Collaboration}%\noaffiliation


\begin{abstract}
The fingering instability of thin-film flow is a significant obstacle to producing uniform and stable films in industrial processes. In the framework of lubrication theory, we investigate the linear stability of a gravity-driven thin film with an inclined contact line using numerical and asymptotic methods. The contact-line singularity is regularized by imposing the Greenspan slip model. As demonstrated by the results, although the contact-line inclination slightly enhances long-wavelength instability, it significantly suppresses the growth of spanwise finite-wavelength perturbations. Moreover, the most unstable wavenumber shifts toward the long-wave region as the inclination angle increases. The inclination of the contact line introduces complex eigenvalues, leading to traveling-wave modes, in contrast to the standing-wave modes observed for horizontal contact lines. We also performed asymptotic analysis in the limit of small wavenumbers, confirming the enhanced long-wave instability. The physical mechanisms underlying these stability characteristics are elucidated, revealing that the effects of contact-line inclination come from the eigenmode distortion and additional streamwise flux.
% \begin{description}
% Secondary publications and information retrieval purposes.
% You may use the \texttt{description} environment to structure your abstract;
% use the optional argument of the \verb+\item+ command to give the category of each item. 
% \end{description}
\end{abstract}

\maketitle

%\tableofcontents

\section{\label{sec:introduction}Introduction}

The generation of thin films driven by a body force is a crucial technique in industrial applications for producing uniform coatings. Its widespread adoption extends to various fields, including microelectronics, optics and other scientific domains \citep{lopesReversiblePolymergelTransition2021,sarkinReviewAntireflectionSelfcleaning2020,liuSpinCoatingSemiconductor2025}. Canonical studies of such force-driven films encompass gravity-driven flow down an inclined plane \citep{huppertFlowInstabilityViscous1982,silviRewettingInclinedSolid1985}, spin coating driven by centrifugal force \citep{schwartzTheoreticalNumericalResults2004,wangAxisymmetricEvolutionThin2025}, Marangoni flows arising from surface tension gradients \citep{cazabatFingeringInstabilityThin1990,brzoskaExponentialGrowthFingering1992}, and flows driven by air jets \citep{moriartyUnsteadySpreadingThin1991a}. Regardless of the particular driving force, this class of problem possesses a common feature: the capillary ridge, a ``bump'' which forms in the vicinity of the contact line, is unstable to spanwise perturbations \citep{troianFingeringInstabilitiesDriven1989}, ultimately leading to a non-uniform film thickness that is incompatible with industrial requirements. The instability eventually appears as a finger-like structure, hence the name ``fingering instability''.

This kind of instability generated by fluid spreading over substrates is accompanied by a moving contact line, which introduces additional challenges to the analysis. In the framework of hydrodynamics, the classical stress singularity arising from the moving contact line can be relieved by one of a hierarchy of models: the slip models \citep{dussanSpreadingLiquidsSolid1979,greenspanMotionSmallViscous1978}, the precursor film model \citep{gennesCapillarityWettingPhenomena2010,voinovWettingLineDynamics2000}, or the diffuse interface model \citep{seppecherMovingContactLines1996}. However, these microscopic models will give rise to a multiscale feature \citep{suiNumericalSimulationsFlows2014}, which often requires additional attention in numerical simulations. Moreover, it is commonly assumed that the film is sufficiently thin to be simplified using lubrication theory \citep{oronLongscaleEvolutionThin1997,crasterDynamicsStabilityThin2009}.

The film flow driven by gravity is a typical and important problem for studying the fingering instability. Relevant studies often focus on the physical problem of a gravity-driven thin film sliding down an inclined plane. In this configuration, when the flow is continuously fed, the base state is steady, making this problem particularly suitable for detailed theoretical and experimental investigations. Early experimental observations of this phenomenon were reported by \citet{huppertFlowInstabilityViscous1982}, who studied viscous films flowing down inclined planes and measured the characteristic wavelength of the resulting fingering patterns. Subsequently, \citet{troianFingeringInstabilitiesDriven1989} performed a linear stability analysis of the advancing front, obtaining the dispersion relation and the most unstable wavelength. The predicted wavelength was found to agree well with the experimental measurements of \citet{huppertFlowInstabilityViscous1982}. Later, \citet{bertozziLinearStabilityTransient1997} demonstrated the existence of a transient growth mechanism in this problem, explaining the inconsistency between experiments \citep{debruynGrowthFingersDriven1992} and the linear theory. Beyond these studies, further explanations for the fingering instability have been proposed. \citet{brennerinstabilitymechanismdriven1993} interpreted the instability by analyzing each term in the lubrication equation, elucidating both the exponential and linear growth of the fingers observed in the experiment \citep{debruynGrowthFingersDriven1992}. Subsequently, \citet{spaidStabilityNewtonianViscoelastic1996} provided a detailed energy analysis of this problem, demonstrating that, during the linear stage, the fingering instability arises primarily from the response of the perturbed free-surface base state to the body force. Specifically, since the thicker regions are more massive, they will be pushed forward more rapidly by the body force, thus causing the fingering instability.
 
With the advancement of related studies, various physical effects influencing the fingering instability have been investigated to explore suitable methods for stabilizing the system. \citet{spaidStabilityNewtonianViscoelastic1996} examined the influence of different contact-line models and non-Newtonian rheology. They demonstrated that the choice of contact-line model had a negligible impact on stability, and the viscoelastic fluid stabilized the system by promoting streamwise momentum transfer. The role of the contact angle was experimentally examined by \citet{silviRewettingInclinedSolid1985}, who pointed out its significant influence on substrate rewetting: small contact angles promote rewetting, which makes it hard to form slender finger-like structures, while larger angles inhibit it. \citet{kondicFlowThinFilms2003} investigated how inhomogeneous surfaces modulate the formation and spacing of the finger structures. In addition, the geometric effects of the substrate have been considered. For instance, \citet{takagiFlowInstabilityThin2010} studied the fingering instability on spherical surfaces, identifying a correlation between critical instability location and the volume of the fluid. Similarly, \citet{balestraFingeringInstabilityCurved2019} performed a non-modal stability analysis on curved cylindrical substrates, obtaining the optimal initial perturbation in relevant parameters. \citet{mayogravitydrivenfingeringsimulations2013} investigated the fingering instability on a vertical cylinder through three-dimensional numerical simulation, revealing the dual effects of spanwise substrate curvature on modulating the instability.

So far, although the fingering instability in free-surface flows has been extensively investigated, effective approaches to modulate the instability characteristics are still limited. Furthermore, prior studies have focused predominantly on configurations where the base flow is perpendicular to the unperturbed contact line, representing a horizontal contact-line configuration, while the effects of contact-line inclination remain unexplored. To explore potential control methods and extend understanding beyond the conventional horizontal-contact-line configuration, this paper therefore focuses on the effects of contact-line inclination on the stability characteristics of the gravity-driven flow. In this configuration, the base flow remains steady and two-dimensional despite the complex three-dimensional flow field induced by the spanwise body force. In particular, we show that the contact-line inclination can significantly suppress the instability for finite-wavelength perturbations, while only slightly enhancing the instability at long-wavelengths. Moreover, in contrast to the horizontal contact line case, both the propagating behavior and the structure of the perturbation field exhibit distinctly different patterns.

This paper is organized as follows. Sec.~\ref{sec:problem formulation} outlines the physical problem and introduces the governing lubrication equations. In Sec.~\ref{sec:base flow}, the base flow formulation is established and its numerical solutions are discussed. Sec.~\ref{sec:Linear stability analysis} is dedicated to the linear stability analysis, presenting both numerical and asymptotic results. The underlying destabilizing or stabilizing mechanisms are explored in Sec.~\ref{sec:Mechanisms} based on these findings. Conclusions are provided in Sec.~\ref{sec:Conclusions}. 

\begin{figure}
  \centering
  \begin{overpic}[width=0.45\textwidth]{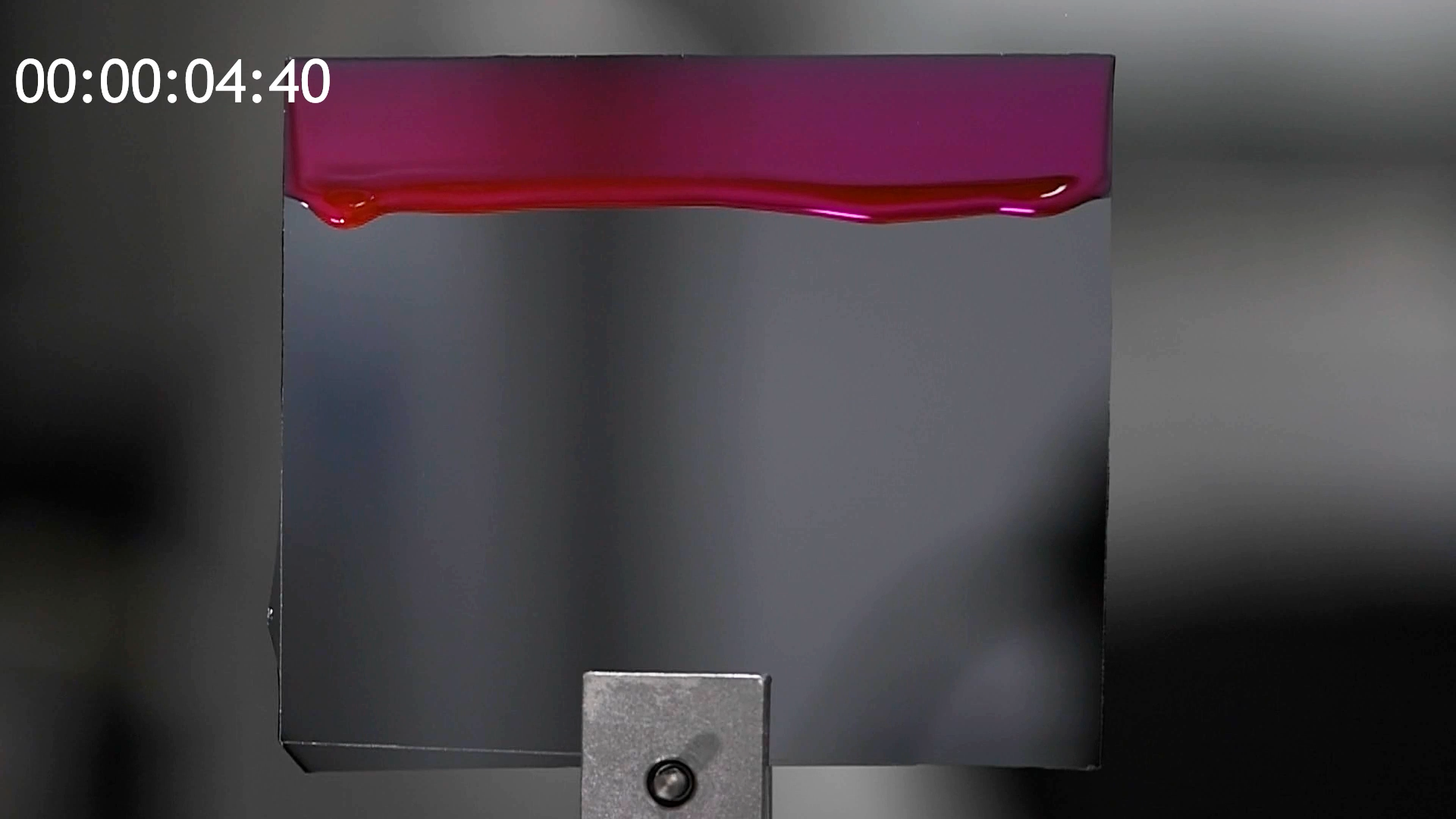}
    \put(-8,50){(a)}
  \end{overpic}
  \qquad
  \begin{overpic}[width=0.45\textwidth]{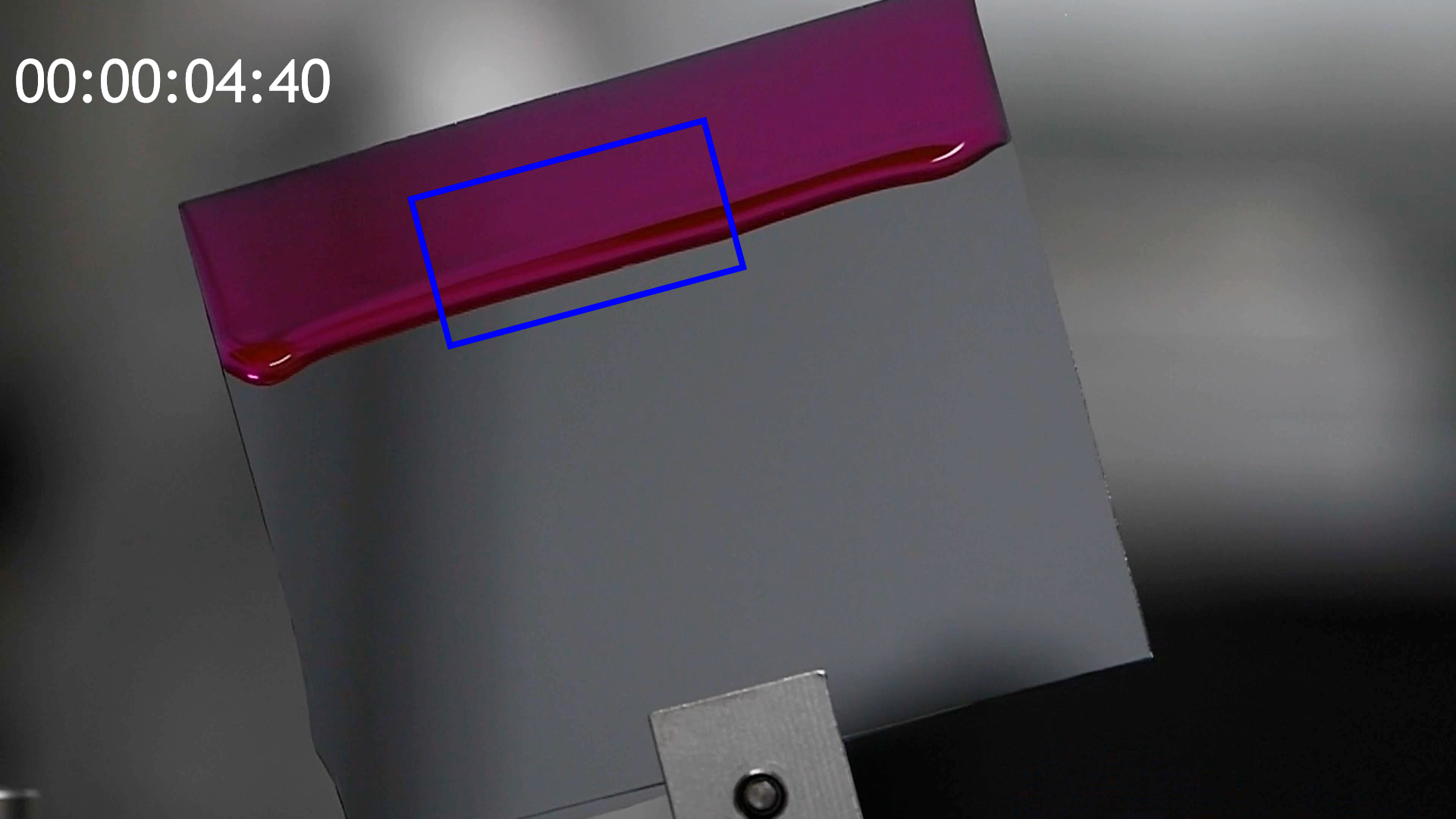}
    \put(-8,50){(b)}
  \end{overpic}
  \\[2pt]
  \begin{overpic}[width=0.45\textwidth]{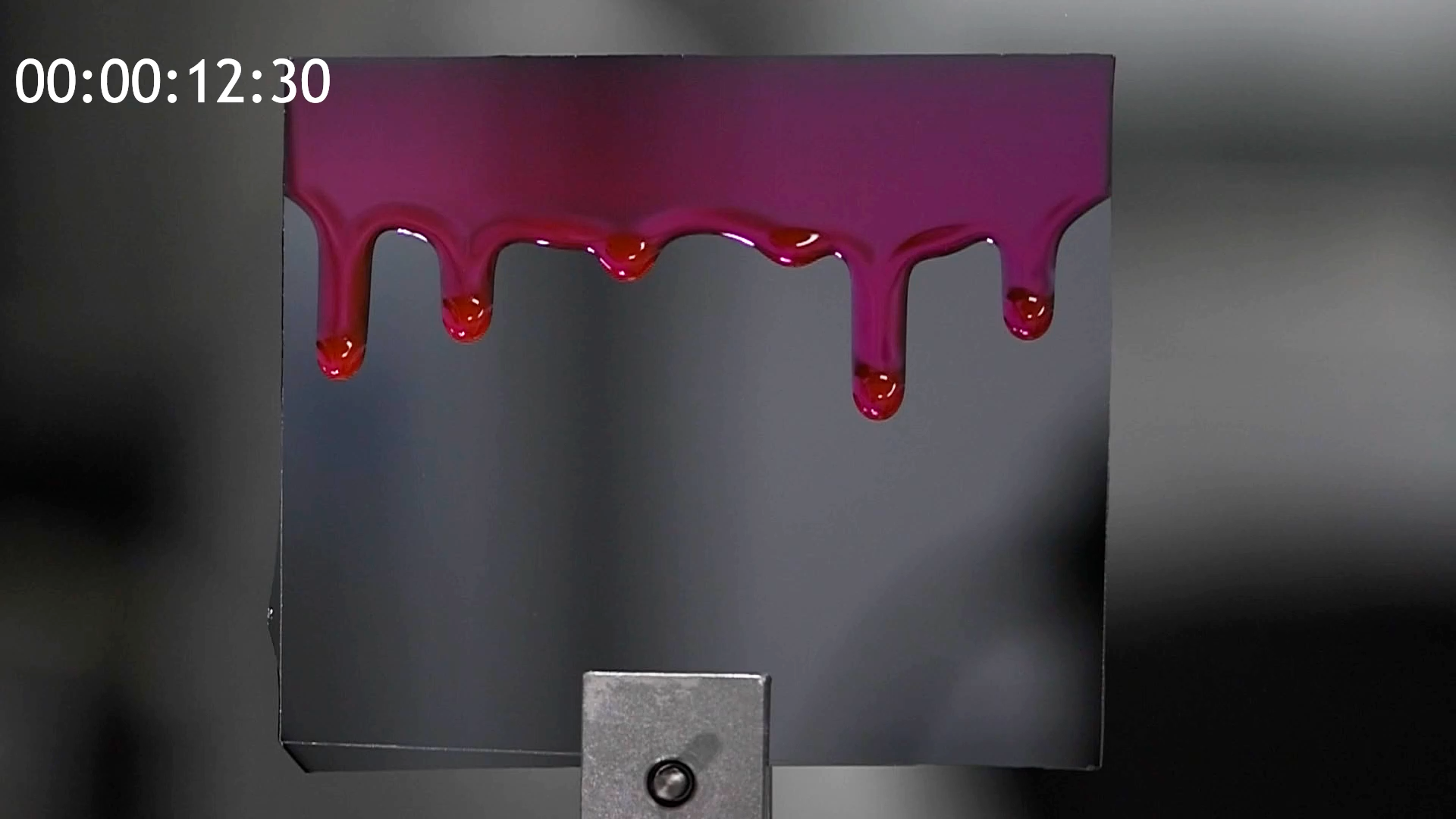}
    \put(-8,50){(c)}
  \end{overpic}
  \qquad
  \begin{overpic}[width=0.45\textwidth]{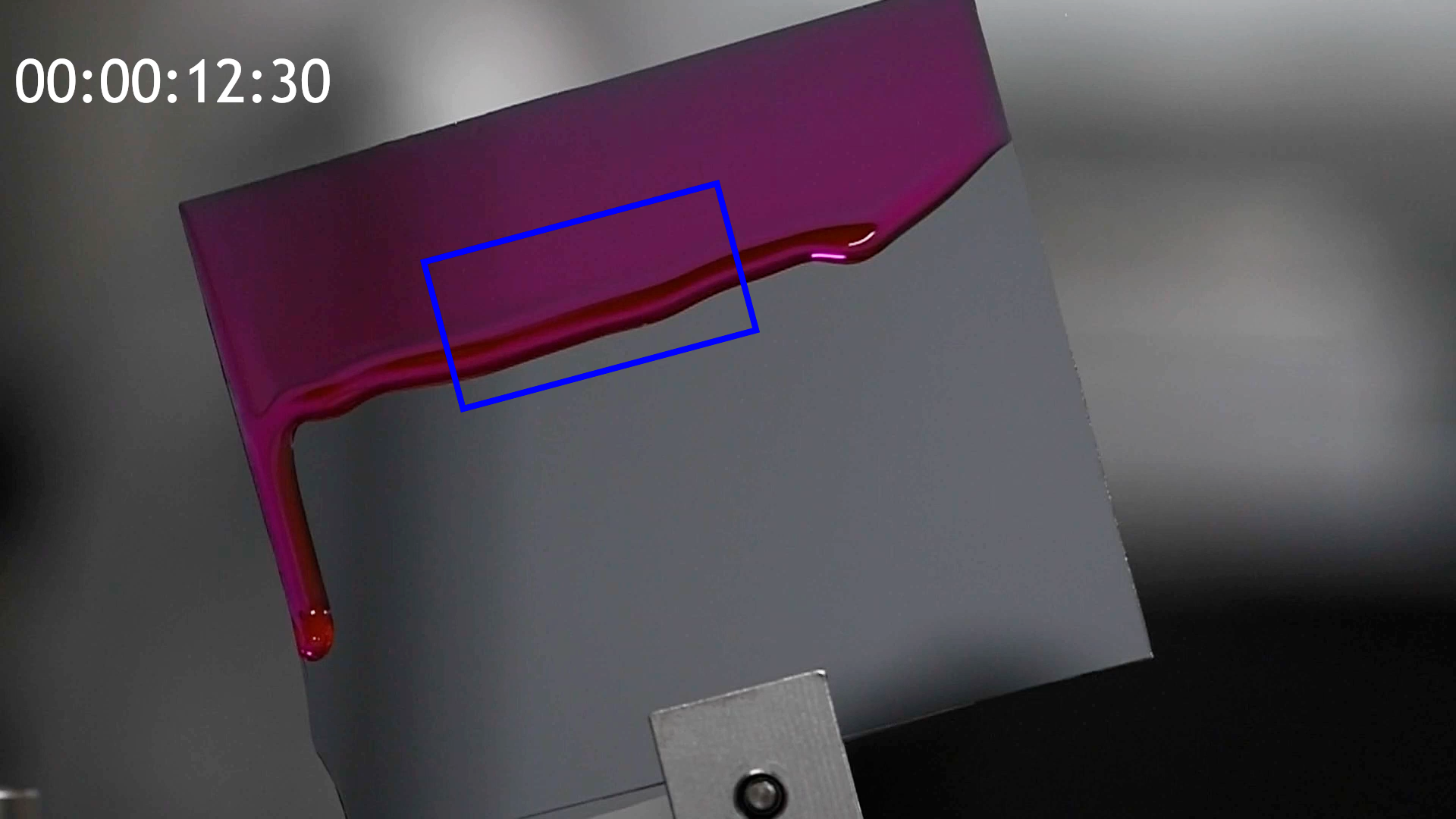}
    \put(-8,50){(d)}
  \end{overpic}
  \caption{Demonstrative experiments illustrating the spreading of a gravity-driven film with (a, c) horizontal and (b, d) inclined contact-line configurations. Snapshots are shown at (a, b) $t = 4.40\text{ s}$ and (c, d) $t = 12.30\text{ s}$. The blue box highlights the region of interest where edge effects are negligible.}
 \label{fig:experimental_sketch}
\end{figure}

\section{\label{sec:problem formulation}Problem formulation}
The physical problem under consideration can be clearly introduced by two simple demonstrative experiments. We first employ a dip-coating technique, wherein a hydrophilically treated silicon wafer ($11\times9.5$ cm) is withdrawn from a bath of dye-doped glycerol-water mixture (10:1 by volume) to deposit a liquid film bounded by a horizontal contact line, with the upper region remaining dry. The silicon wafer is then inverted to allow gravity-driven spreading, yielding the early- and late-stage images of film spreading shown in Fig.~\ref{fig:experimental_sketch}(a, c). As expected, the traditional fingering instability occurs over the whole front of the film. In the other experiment, the wafer is rotated by a prescribed angle after the inversion to introduce an inclined contact line. As shown in Fig.~\ref{fig:experimental_sketch}(b, d), no significant fingering structures are observed except near the edges. The central region of the inclined contact line (roughly enclosed by the blue box) is less prone to instability, while the contact line itself moves downward at an approximately constant inclination angle. The edge effects are expected to be relatively negligible in the central region, which is the focus of this work.

\begin{figure}
  \centerline{\includegraphics[scale=.45]{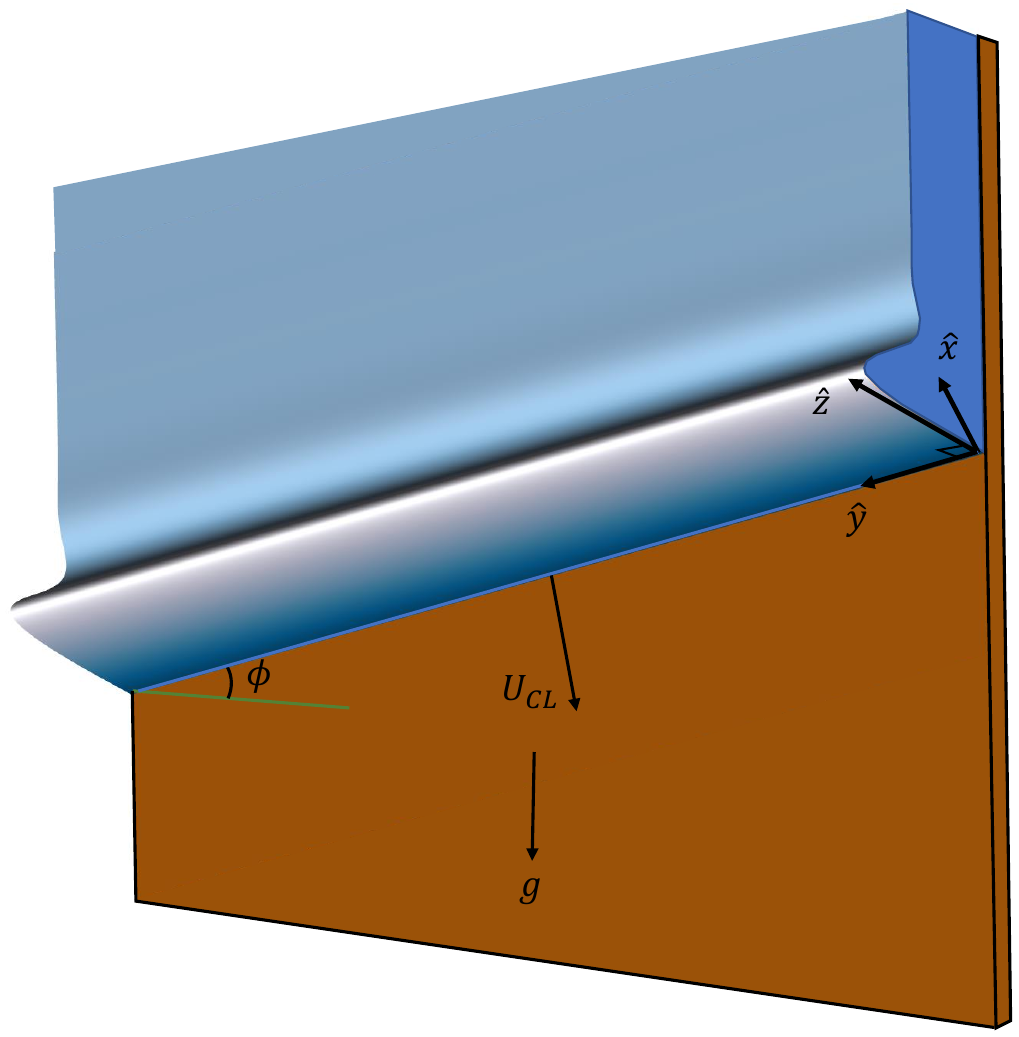}}
  \caption{Sketch of a gravity-driven thin film with an inclined contact line.}
\label{fig:physical_model_2}
\end{figure}

To effectively model this blue-box region, we consider a gravity-driven thin liquid layer with an inclined contact line, as sketched in Fig.~\ref{fig:physical_model_2}. The plate is assumed to be unbounded in both the streamwise and spanwise directions so that edge effects are negligible. Without perturbations, the flow is assumed to be fully developed with the contact line propagating downward at a constant normal speed $U_{\mathrm{CL}}$. It is convenient to employ a moving coordinate system $(\hat{{x}}, \hat{{y}}, \hat{{z}})$ fixed to the contact line, in which the base flow is steady and the plate moves along $\hat{x}$-axis with a speed $U_{\mathrm{CL}}$. The inclination angle of the contact line is denoted by $\phi$, the contact angle by $\hat\theta_e$, the gravitational acceleration by $g$, and the position of the free surface by $\hat{z}=\hat{h}(\hat{{x}}, \hat{{y}}, t)$, where $t$ is time. The fluid density, dynamic viscosity, and surface tension coefficient are denoted by $\rho$, $\mu$ and $\sigma$, respectively.

It is known that the presence of the moving contact line gives rise to a stress singularity \citep{huhHydrodynamicModelSteady1971,dussanv.MotionFluidfluidInterface1974}, which should be regularized by introducing a microscopic contact-line model. Previous studies \citep{sibleyComparisonSlipDisjoining2015,spaidStabilityNewtonianViscoelastic1996} have established that the macroscopic behavior is insensitive to the particular choice of microscopic models. The most widely used contact-line models are slip models, which are introduced through the slip boundary condition at the wall. Specifically, the Navier slip model adopts a constant slip length $\lambda_s$, with the slip boundary condition expressed as
\begin{equation}
    \hat{\boldsymbol{u}}=\boldsymbol{U}_{\text{w}}+\frac{\lambda_s}{\mu}\boldsymbol{T}_\text{w},
\end{equation}
where $\hat{\boldsymbol{u}}$ is the fluid velocity at the wall, $\boldsymbol{U}_{\text{w}}$ is the wall velocity, and $\boldsymbol{T}_{\text{w}}$ represents the wall shear stress. Alternatively, the Greenspan slip model \citep{greenspanMotionSmallViscous1978} employs a variable slip length that depends on the local film thickness. In this work, we adopt the Greenspan slip model, which is given by
\begin{equation}
    \hat{\boldsymbol{u}}=\boldsymbol{U}_{\text{w}}+\frac{\hat{\alpha}}{\hat{h}} \frac{\boldsymbol{T}_{\text{w}}}{\mu},
\label{eq:slip_condition}
\end{equation}
where $\hat{\alpha}$ is the Greenspan slip parameter with a dimension of squared length. Compared to the Navier slip model, the Greenspan model is preferred here because it fully eliminates the contact-line singularity and is more appropriate for numerically solving the lubrication equation.

The present work concentrates on the situation of small contact angles and negligible inertia effects such that the dynamics can be well described by lubrication theory \citep{oronLongscaleEvolutionThin1997}. In order to render the problem dimensionless, we scale the variables as
\begin{equation}
      {x} =\frac{\hat{{x}}}{L},\quad {y}=\frac{\hat{{y}}}{L},\quad {z}=\frac{\hat{{z}}}{H},\quad h=\frac{\hat{h}}{H},\quad t=\frac{3\mu L^4}{\sigma H^{3}}\tau,\quad \alpha=\frac{\hat{\alpha}}{H^2} ,\quad \theta_e = \frac{L}{H}\hat{\theta}_e,
      \label{dimensionless_var}
\end{equation}
where $L= \left( \sigma H / (\rho g \cos \phi) \right)^{1/3}$ is the modified capillary length and $H$ denotes the uniform film thickness far from the contact line. This yields the dimensionless lubrication equation
\begin{equation}
      \frac{\partial h}{\partial \tau}+\nabla\cdot\boldsymbol{Q}=0,
      \label{dimensionless-lubri-eq}
\end{equation}
where the flux vector in the current coordinate system is given by
\begin{equation}
    \boldsymbol{Q}= h(h^{2}+\alpha)\nabla ({\nabla^2h}-{x}+{y}\tan\phi)+\delta h \boldsymbol{\mathrm{e}}_{x}.
    \label{dimensionless-flux}
\end{equation}
Here, $\nabla\equiv(\partial/\partial{x},\partial/\partial{y})$, $\boldsymbol{\mathrm{e}}_{{x}}$ represents the unit vector in the ${x}$ direction, and
\begin{equation}
      \delta =\frac{3\mu U_{\mathrm{CL}}}{\rho g H^{2}\cos\phi}
      \label{eq:delta}
\end{equation}
is the dimensionless mean velocity of the contact line. Specifically, the first three individual contributions to the flux $\boldsymbol{Q}$, which are proportional to $h(h^2+\alpha)$, arise from capillary forces, the streamwise body force, and the spanwise body force, respectively. The final contribution, $\delta h \boldsymbol{\mathrm{e}_x}$, stems from the substrate motion within the adopted reference frame. A detailed derivation of the lubrication equation can be found in \cite{oronLongscaleEvolutionThin1997}.
 
The lubrication equation is supplemented by the following boundary conditions. At the contact line, the film thickness satisfies
\begin{equation}
      h=0,\quad \boldsymbol{n}\cdot\nabla h=-\theta_e,
      \label{contact_line_boundary}
\end{equation}
where $\boldsymbol{n}$ denotes the outward unit normal vector at the contact line. In the far field, the film thickness is assumed to be uniform, so that
\begin{equation}
      h\to 1 \quad \mbox{as} \quad {x}\to\infty.
      \label{far_field_boundary}
\end{equation}
Since the position of the perturbed contact line remains unknown, a further condition is needed to close the problem. Supposing the contact line is given by ${x}={x}_c({y},\tau)$, we enforce a kinematic boundary condition at the contact line,
\begin{align}
      \frac{\partial{x}_c}{\partial\tau}{\boldsymbol{\mathrm{e}}_{{x}}\cdot\boldsymbol{n}} = \boldsymbol{\bar{u}}\cdot \boldsymbol{n},
      \label{BC_kinematic_condition}
\end{align}
where $\boldsymbol{\bar{u}}=\boldsymbol{Q}/h$ denotes the depth-averaged velocity. This condition guarantees that the contact line is a material line.

\section{\label{sec:base flow}Base flow}
The governing equation~Eq.~\eqref{dimensionless-lubri-eq} together with the boundary conditions admits a steady solution of the form $h=h_0({x})$, which serves as the base state of present interest. After one integration,~Eq.~\eqref{dimensionless-lubri-eq} reduces to
\begin{equation}
      h_{0{x}{x}{x}} = 1 - \frac{\delta }{h_{0}^{2} + \alpha}.
      \label{base_flow_eq_pre}
\end{equation} 
The corresponding contact line is straight and represented by ${x}=0$, where the boundary conditions are
\begin{equation}
      h_0=0,\quad h_{0{x}}=\theta_e.
      \label{base_flow_cl_boundary}
\end{equation}
In the far field, we have
\begin{equation}
      h_0 \to 1 \quad \mbox{as} \quad {x}\to\infty.
      \label{base_flow_far_boundary}
\end{equation}
Using this far-field condition, it can be obtained from Eq.~\eqref{base_flow_eq_pre} that
\begin{equation}
      \delta = 1 + \alpha,
      \label{eq:delta_expression}
\end{equation}
and hence the contact line speed is determined by the far-field film thickness. Substituting~Eq.~\eqref{eq:delta_expression} into~Eq.~\eqref{base_flow_eq_pre} yields the final equation governing the base flow:
\begin{equation}
      h_{0{x}{x}{x}} = \frac{h_0^2 - 1}{h_{0}^{2} + \alpha}.
      \label{base_flow_eq}
\end{equation}

The base flow equation Eq.~\eqref{base_flow_eq} and boundary conditions Eqs.~\eqref{base_flow_cl_boundary} and \eqref{base_flow_far_boundary} are independent of $\phi$ and formally identical to those for the case with a non-inclined contact line \citep{Tuck1990,spaidStabilityNewtonianViscoelastic1996}. Therefore, the basic film profiles remain unchanged with the scalings in Eq.~\eqref{dimensionless_var}. We note, however, that the configuration of the inclined contact line still represents a distinct flow state. Specifically, the base flow is two-dimensional for a horizontal contact line, while the presence of the inclined contact line gives rise to a three-dimensional flow field \citep[see also][for the case of receding contact lines with inclination]{gaoDewettingFilmsInclined2015}. From Eq.~\eqref{dimensionless-flux}, it follows directly that the flux in the ${x}$ direction vanishes, i.e., $Q_{\perp}=0$, while the flux in the ${y}$ direction is
\begin{equation}
    Q_\parallel=h_0(h_0^2+\alpha)\tan\phi,
    \label{base_flow_eta-flux}
\end{equation}
which is finite when the contact line is inclined. This tangential flux $Q_\parallel$ is induced by the spanwise component of gravity and gives rise to liquid accumulation on the lower side of the contact line for a plate of finite width, as illustrated in Fig.~\ref{fig:experimental_sketch}(d). However, such accumulation occurs outside the physical region of interest, where the edge effects are negligible. Notably, in this problem, the tangential flux approaches a uniform value $(1+\alpha)\tan\phi$ in the far field, in contrast to the vanishing flux reported for dewetting films with an inclined contact line \citep{gaoDewettingFilmsInclined2015}. This discrepancy arises from the adoption of different reference frames. In our current coordinate system, the substrate motion is perpendicular to the contact line and thus cannot cancel out the transverse flux induced by gravity. By contrast, in the coordinate system adopted by \citet{gaoDewettingFilmsInclined2015}, the substrate motion is vertical, canceling both the streamwise and spanwise components of the gravity-driven flux in the far field.

\begin{figure}
  \centering
  \begin{overpic}[width=0.55\textwidth]{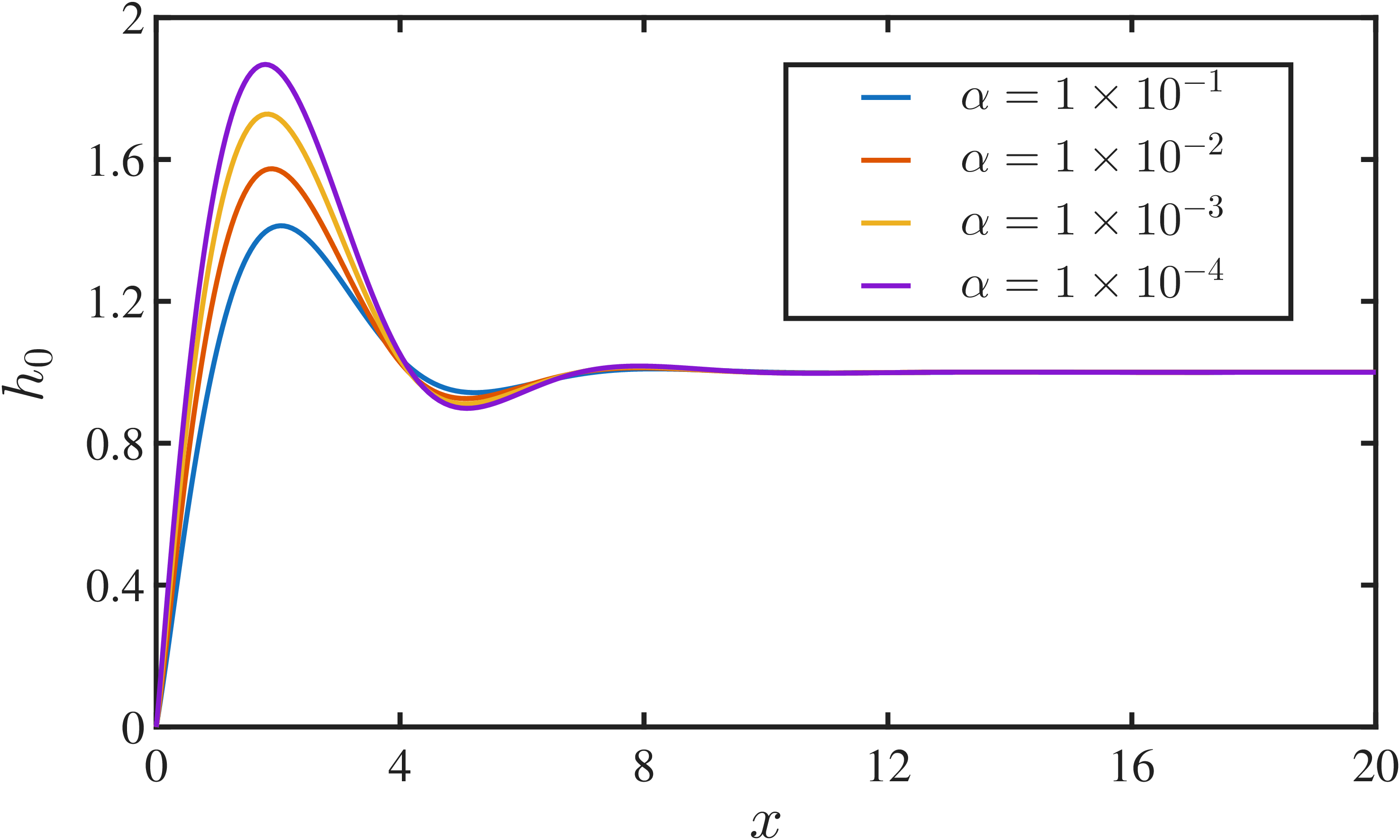}
    \put(-5,55){(a)}
  \end{overpic}
  \\[10pt]
  \begin{overpic}[width=0.55\textwidth]{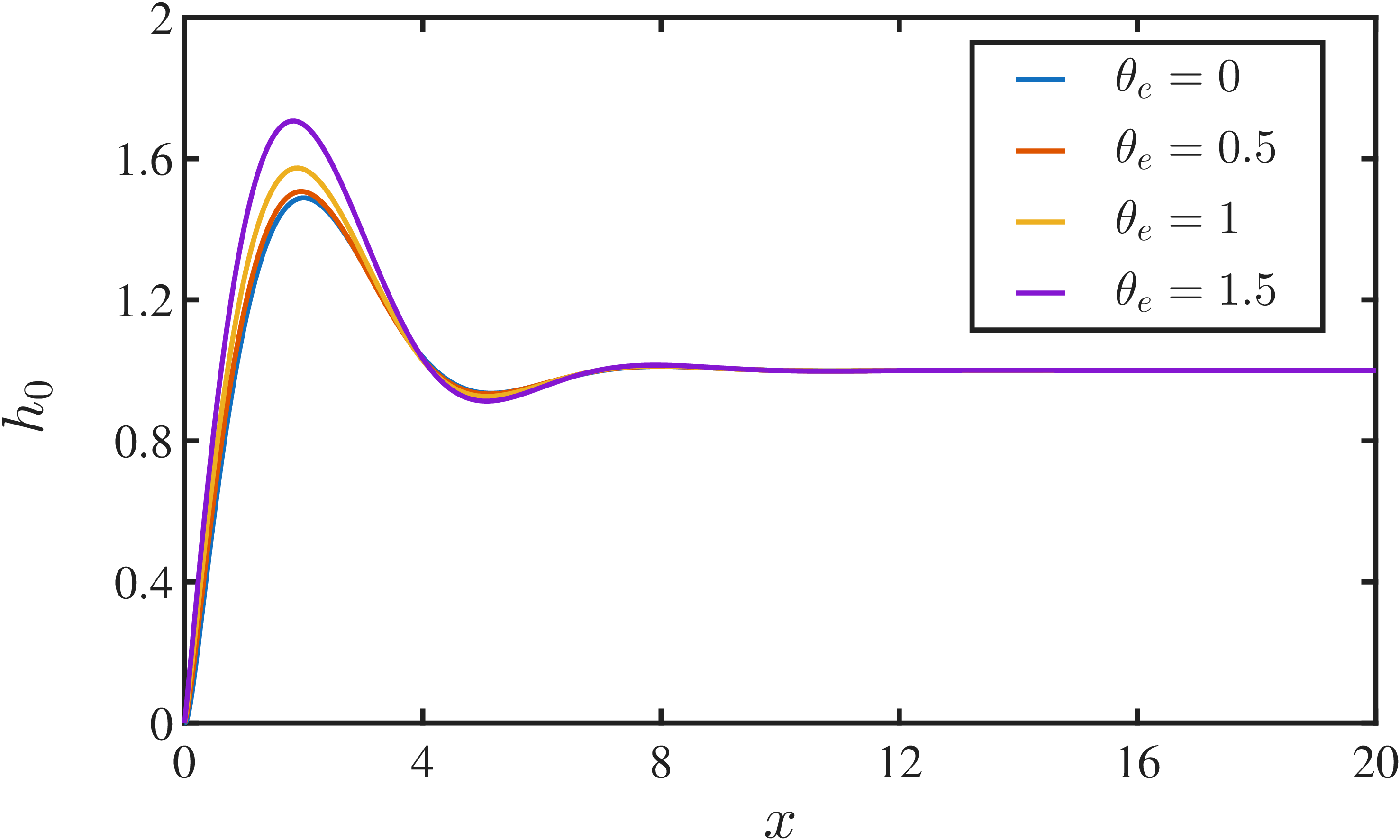}
    \put(-5,55){(b)}
  \end{overpic}
  \caption{Base film profiles $h_0({x})$ for representative values of (a) the slip parameter $\alpha$ with $\theta_e = 1$ and (b) the equilibrium contact angle $\theta_e$ with $\alpha = 10^{-2}$.}
  \label{fig:base_flow_profile}
\end{figure}

The base film profile is governed by two dimensionless parameters, i.e., $\alpha$ and $\theta_e$, and has been investigated by \citet{spaidStabilityNewtonianViscoelastic1996}. For completeness, Fig.~\ref{fig:base_flow_profile} illustrates the base-flow profiles for representative values of these parameters. The film is characterized by a capillary ridge near the contact line, which is the primary cause of fingering instability \citep{troianFingeringInstabilitiesDriven1989}. As shown in Fig.~\ref{fig:base_flow_profile}(a), a reduction in the slip parameter leads to a marked increase in the maximum height of the capillary ridge. Regarding the influence of the contact angle, Fig.~\ref{fig:base_flow_profile}(b) reveals that for small contact angles (specifically $\theta_e < 0.5$), the profiles remain insensitive to variations in $\theta_e$. Conversely, for larger contact angles, an increase of $\theta_e$ results in a more prominent capillary ridge. 

\section{\label{sec:Linear stability analysis}Linear stability analysis}

\subsection{Linearized perturbation equation}
In the framework of the linear stability theory, the film thickness can be decomposed as
\begin{equation}
      h({x} ,{y} ,\tau ) = h_{0}({x} ) + \epsilon h_{1}({x},{y},\tau),
      \label{perturbation_form}
\end{equation}
where $\epsilon \ll 1$ characterizes the perturbation amplitude. Since the base state is independent of ${y}$, the perturbation can be expressed in normal-mode form
\begin{equation}
      h_1({x},{y},\tau) = \mbox{Re}\left\{G({x})e^{{i}q{y} + \beta\tau}\right\},
      \label{regular_mode}
\end{equation}
where $G({x})$ is the eigenfunction, $q$ is the spanwise wavenumber and $\beta$ is the eigenvalue with its real part representing the temporal growth rate. Substituting Eqs.~\eqref{perturbation_form} and \eqref{regular_mode} into the lubrication equation \eqref{dimensionless-lubri-eq} and linearizing with respect to $\epsilon$ yields the stability equation
\begin{align}
      \beta G &+ (1+\alpha)G_{{x}} 
      + h_{0}(h_{0}^{2}+\alpha)(q^{4}G-q^{2}G_{{x}{x}})+ {i}qG(3h_{0}^{2}+\alpha)\tan\phi \notag    \\
      &+ \frac{d}{d{x}}\big[h_{0}(h_{0}^{2}+\alpha)(G_{{x}{x}{x}}-q^{2}G_{{x}})+(h_{0{x}{x}{x}}-1)      (3h_{0}^{2}+\alpha)G\big] = 0.
      \label{Linearized_perturbation_eq_old}
\end{align}
Using Eq.~\eqref{base_flow_eq}, this equation can be slightly simplified to
\begin{align}
      \beta G &
      + h_{0}(h_{0}^{2}+\alpha)(q^{4}G-q^{2}G_{{x}{x}})+ {i}qG(3h_{0}^{2}+\alpha)\tan\phi \notag    \\
      &+ \frac{d}{d{x}}\big[h_{0}(h_{0}^{2}+\alpha)(G_{{x}{x}{x}}-q^{2}G_{{x}})+2h_0^2(h_{0{x}{x}{x}}-1)      G\big] = 0.
      \label{Linearized_perturbation_eq}
\end{align}

For this eigenvalue problem, the following boundary conditions should be satisfied. In the far field, the perturbation decays:
\begin{equation}
      G \to 0, \quad G_{{x}} \to 0, \quad \mbox{as}\quad{x} \to \infty.
      \label{Linearized_perturbation_far_boundary}
\end{equation}
The boundary conditions at the contact line can be obtained by linearizing Eqs.~\eqref{contact_line_boundary} and \eqref{BC_kinematic_condition} around ${x}=0$ and are derived with details provided in Appendix~\ref{boundary_condition}. One of the boundary conditions reads
\begin{equation}
      G - G_{{x}}\frac{h_{0{x}}}{h_{0{x}{x}}} = 0\quad \mbox{at}\quad {x}=0,
      \label{eq:G0BC}
\end{equation}
which reduces to $G=0$ for the case of complete wetting. The form of the last condition at ${x}=0$ differs for partial and complete wetting, and is written as
\begin{equation}
      \beta G + \alpha \left[h_{0{x}}(G_{{x}{x}{x}} - q^2 G_{{x}}) + {i}qG\tan\phi   \right] = 0 \quad\mbox{when}\quad \theta_e \neq 0,
      \label{Linearized_perturbation_cl_boundary_partial}
\end{equation}
and 
\begin{gather}
      %G = 0, \\
      \beta G_{{x}} + \alpha \left[h_{0{x}{x}}(G_{{x}{x}{x}} - q^2 G_{{x}}) + {i}q G_{{x}}\tan\phi  \right] = 0\quad\mbox{when}\quad \theta_e = 0.
      \label{Linearized_perturbation_cl_boundary_complete}
\end{gather}
It can be verified that the situation of horizontal contact lines and partial wetting considered by \citet{spaidStabilityNewtonianViscoelastic1996} can be recovered from Eqs.~\eqref{Linearized_perturbation_eq}--\eqref{Linearized_perturbation_cl_boundary_partial} by setting $\phi=0$.

\subsection{Numerical results of the eigenvalue problem}\label{subsec:numerical results}
\begin{figure}
  \centering
  \begin{overpic}[width=0.55\textwidth]{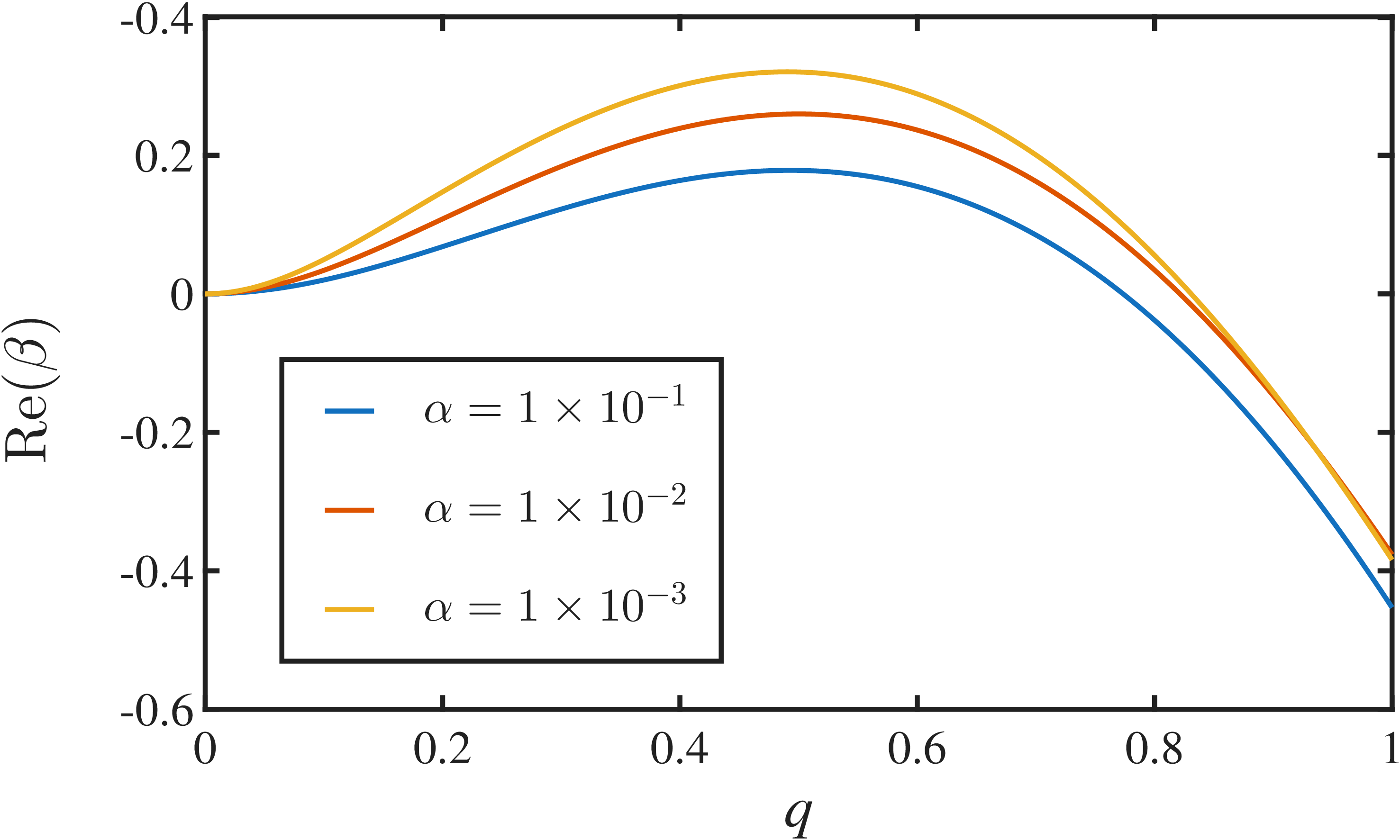}
    \put(-5,55){(a)}
  \end{overpic}
\\[10pt]
  \begin{overpic}[width=0.55\textwidth]{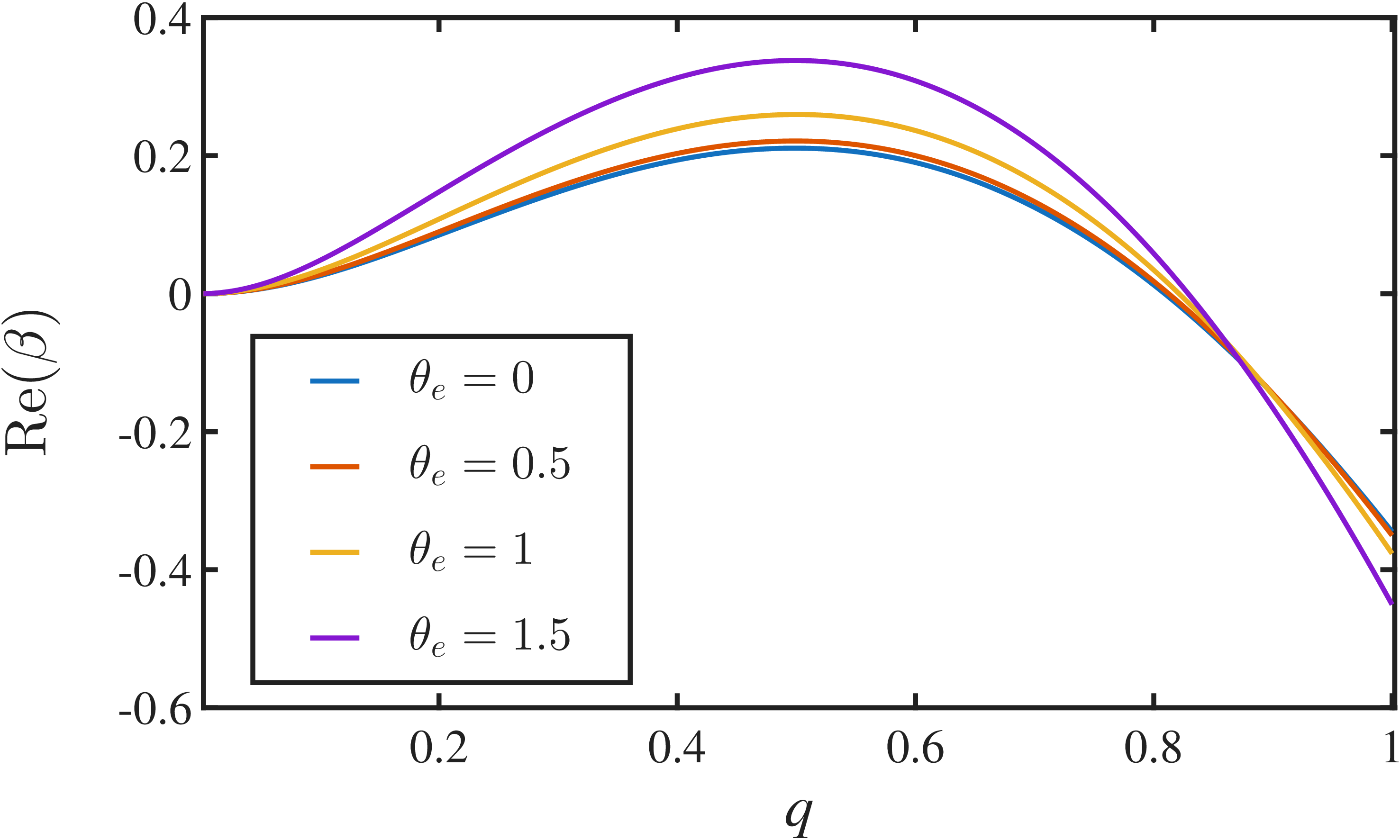}
    \put(-5,55){(b)}
  \end{overpic}
  \caption{Dispersion relations illustrating the growth rate $\mathrm{Re}(\beta)$ as a function of the wavenumber $q$ for $\phi=0$: (a) effect of the slip parameter $\alpha$ at $\theta_e = 1$; (b) effect of the equilibrium contact angle $\theta_e$ at $\alpha = 10^{-2}$.}
  \label{fig:dispersion-curve_alpha-thetae}
\end{figure}

The eigenvalue problem is characterized by three parameters, i.e., $\alpha$, $\phi$ and $\theta_e$, with the effect of $\phi$ being the primary focus of interest. 
We discretized the linearized system using a spectral collocation method with the Chebyshev differentiation matrices provided by \citet{schmidStabilityTransitionShear2001}. The details of this discretization method are briefly discussed in Appendix~\ref{numerical_method}.
% We discretized the Linearized system using the Chebyshev spectral collocation method with the differentiation matrices provided by \citet{schmidStabilityTransitionShear2001}. 
This approach can locally refine the mesh in the vicinity of the contact line and provide spectral accuracy of the solutions. It can be verified that Eqs.~\eqref{Linearized_perturbation_cl_boundary_partial} and \eqref{Linearized_perturbation_cl_boundary_complete} are actually the limit of the governing equation Eq.~\eqref{Linearized_perturbation_eq} and its derivative as ${x}\to0$. Therefore, as confirmed numerically, these conditions may be dropped in the collocation method.

We first present the stability results of the film with a horizontal contact line. Consistent with \citet{troianFingeringInstabilitiesDriven1989} and \citet{spaidStabilityNewtonianViscoelastic1996}, it is found that the eigenvalue of the dominant mode is real, and hence the corresponding perturbations behave as standing waves. Fig.~\ref{fig:dispersion-curve_alpha-thetae} shows the growth rate of the dominant mode as a function of the wavenumber $q$ for typical values of $\alpha$ and $\theta_e$ at $\phi=0$. For all parameters considered, there exists a cutoff wavenumber, below which the eigenvalue is positive, rendering the film flow unstable. The maximum $\beta$ occurs at approximately $q=0.5$, which is insensitive to $\alpha$ or $\theta_e$.

\begin{figure}
    \centering
    \includegraphics[width=0.55\textwidth]{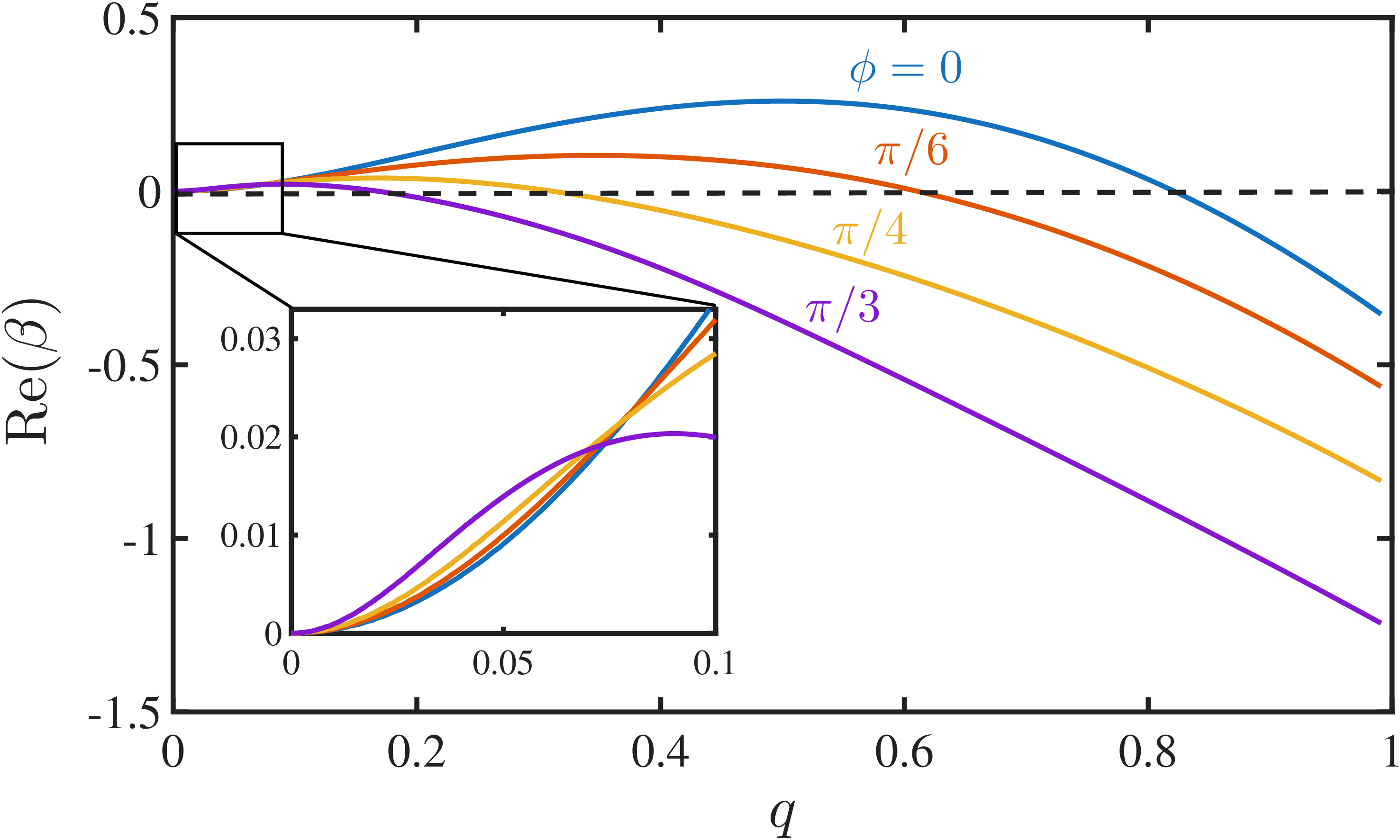}
    \caption{Dispersion curves illustrating the growth rate $\mathrm{Re}(\beta)$ as a function of the wavenumber $q$ for representative inclination angles $\phi$, with $\alpha = 10^{-2}$ and $\theta_e = 1$. The inset shows the magnification of the long-wave region.}
    \label{fig:dispersion_inclination_angle}
\end{figure}

When the contact line is inclined, our numerical results show that the eigenvalue of the dominant mode is complex. Fig.~\ref{fig:dispersion_inclination_angle} presents the growth rate $\mathrm{Re}(\beta)$ of the dominant mode under representative inclination angles $\phi$ at $\alpha=10^{-2}$ and $\theta_e=1$. It is observed that as the inclination angle $\phi$ increases, the growth rate of perturbations in the finite-wavelength region is significantly reduced. Moreover, one can also observe a reduction of the cutoff wavenumber, beyond which the perturbations are stable; in other words, the bandwidth of the unstable modes is reduced as well. In the long-wave region, however, a slight elevation of the growth rates occurs, as highlighted in the inset. Therefore, a weak instability of the film is still present.

\begin{figure}
  \centering
  \begin{overpic}[width=0.45\textwidth]{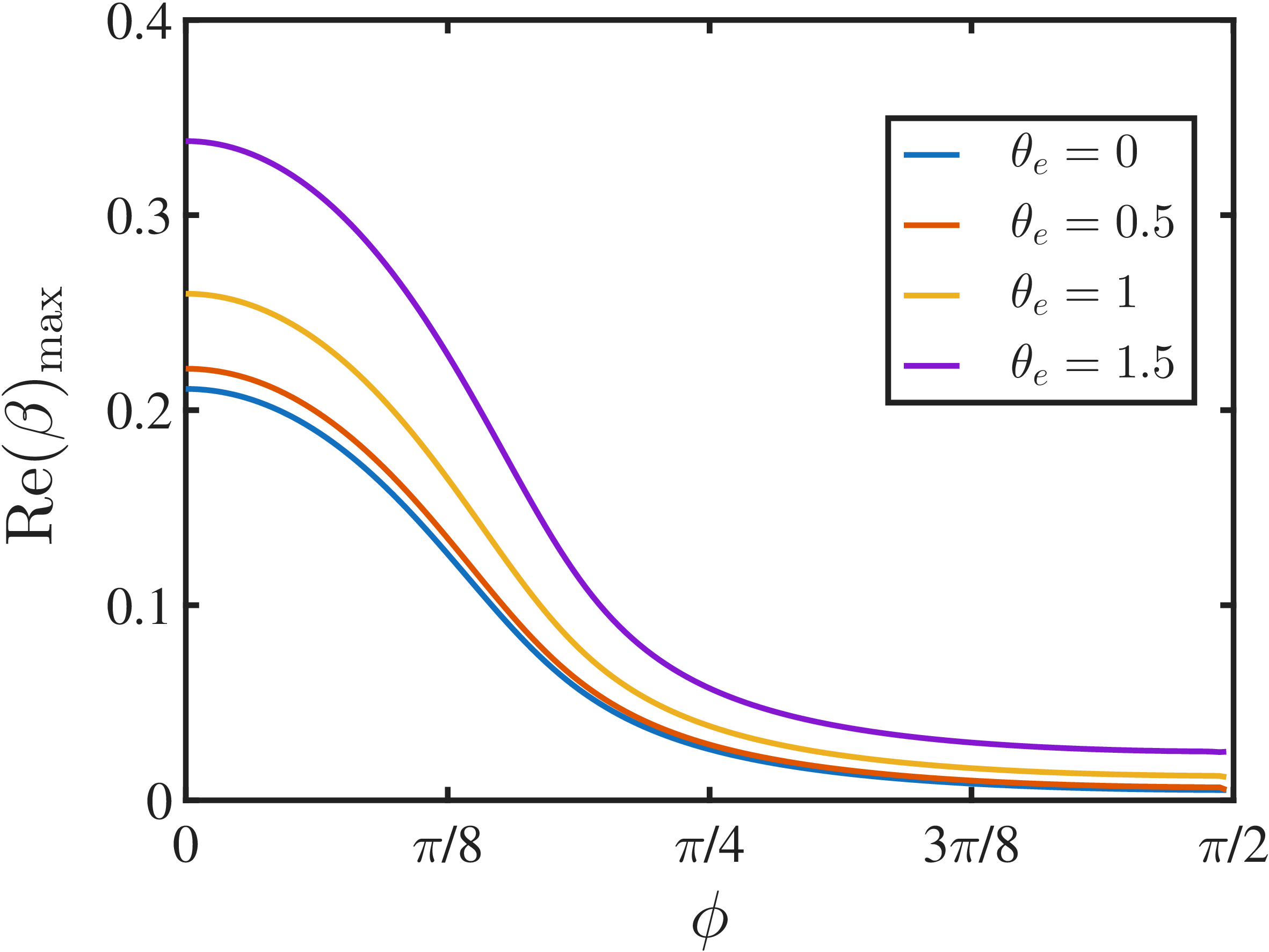}
    \put(-1,70){(a)}
  \end{overpic}
  \quad
  \begin{overpic}[width=0.45\textwidth]{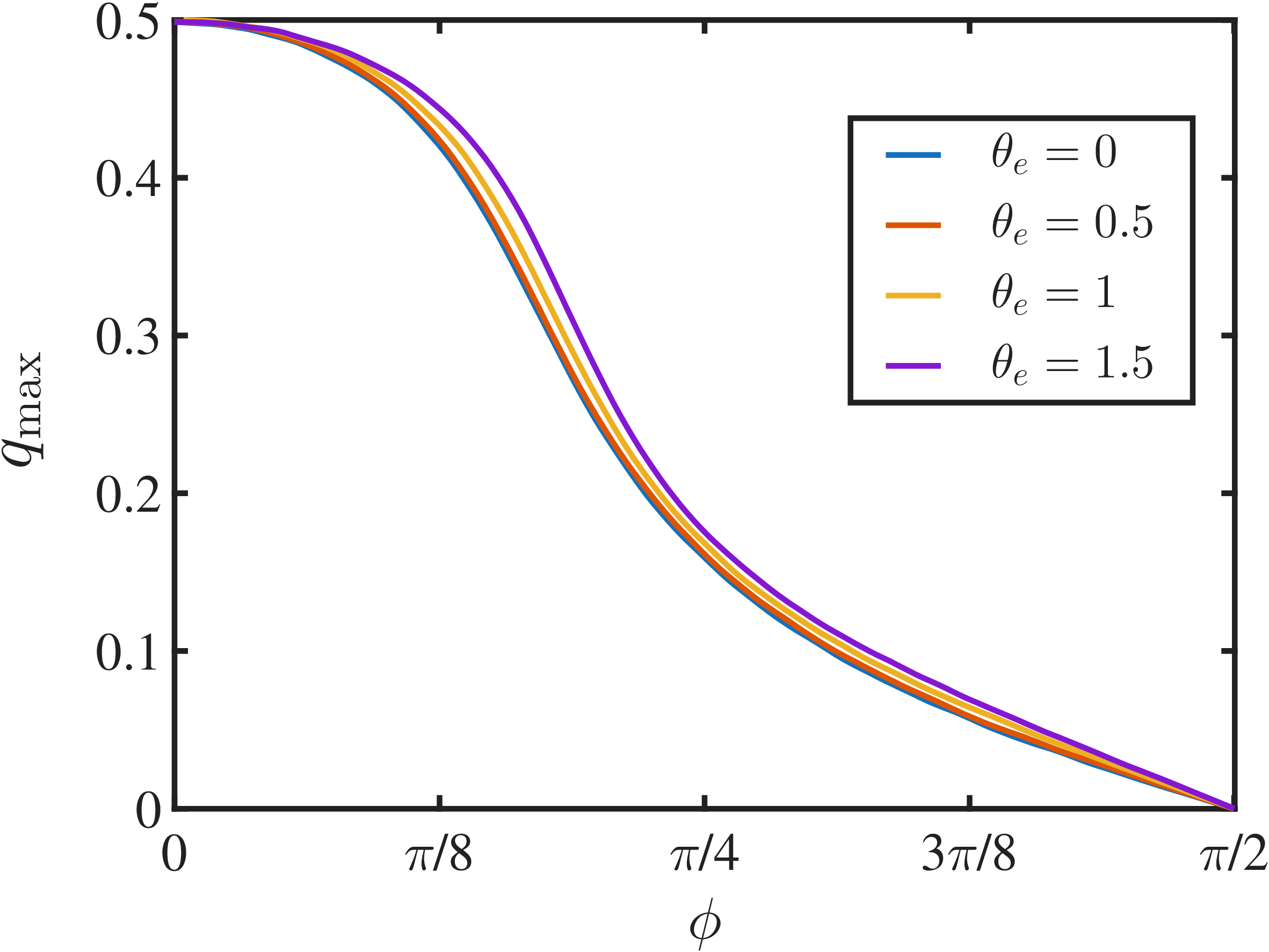}
    \put(-1,70){(b)}
  \end{overpic}
  \caption{Dependence of (a) the maximum growth rate $\mathrm{Re}(\beta)_{\max}$ and (b) the corresponding wavenumber $q_{\max}$ as functions of $\phi$ for $\alpha = 10^{-2}$ and representative contact angles $\theta_e$.}
  \label{result:amplified_wavelength}
\end{figure}

The maximum growth rate $\mathrm{Re}(\beta)_{\max}$ and the corresponding wavenumber $q_{\max}$ are plotted as functions of $\phi$ in Fig.~\ref{result:amplified_wavelength}, where the influence of the contact angle $\theta_e$ is also investigated. Consistent with Fig.~\ref{fig:dispersion_inclination_angle}, both $\mathrm{Re}(\beta)_{\max}$ and $q_{\max}$ decrease markedly with increasing $\phi$ for all the values of $\theta_e$ considered, indicating that the instability is significantly weakened in the presence of the inclination. While the contact angle $\theta_e$ plays an important role in the growth rate, it only weakly influences the wavenumber of the most amplified mode, as shown in Fig.~\ref{result:amplified_wavelength}(b). 

\begin{figure}
    \centering
    \begin{overpic}[width=0.45\textwidth]{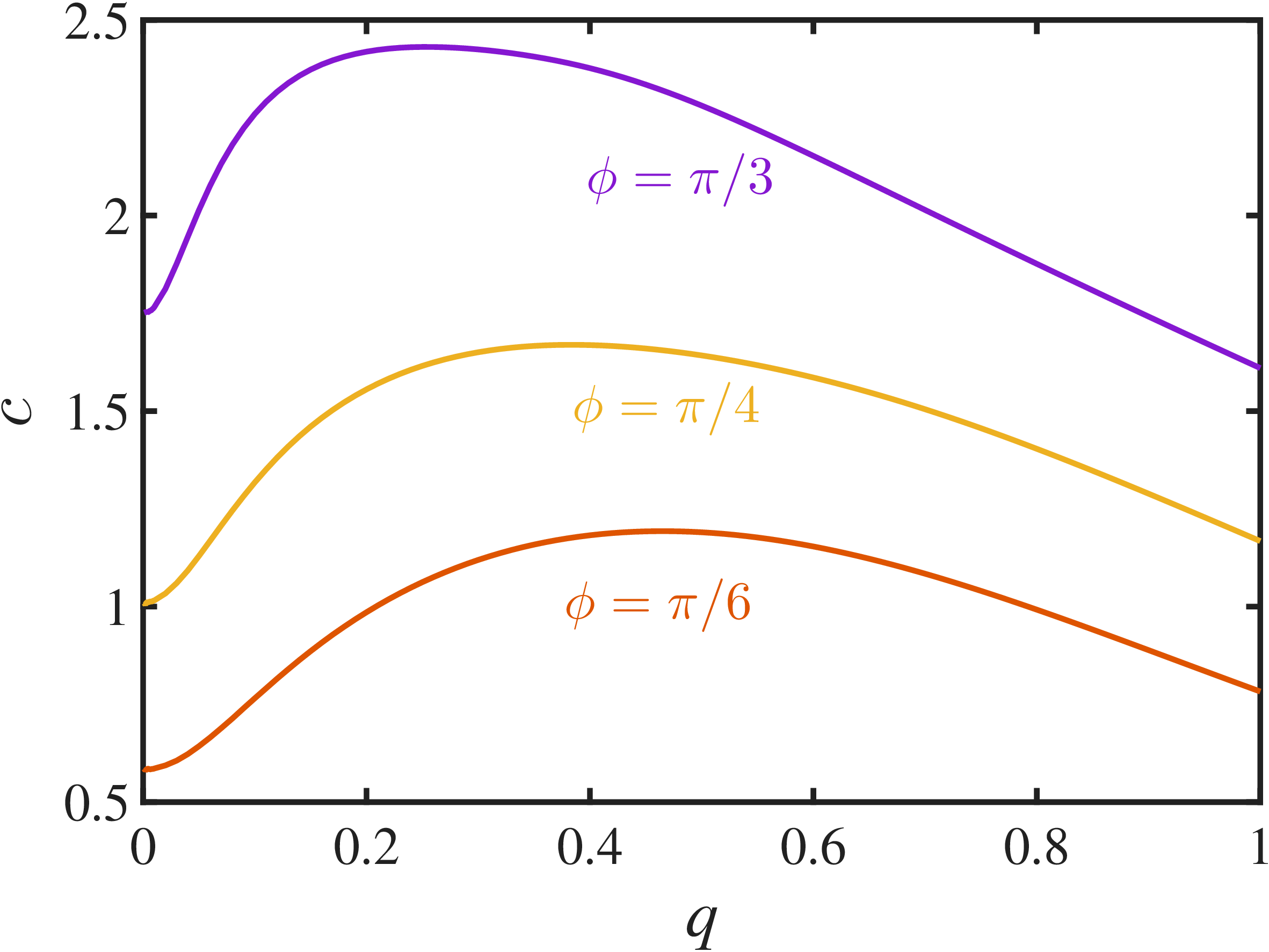}
    \put(-1,70){(a)}
    \end{overpic}
   \quad
    \begin{overpic}[width=0.45\textwidth]{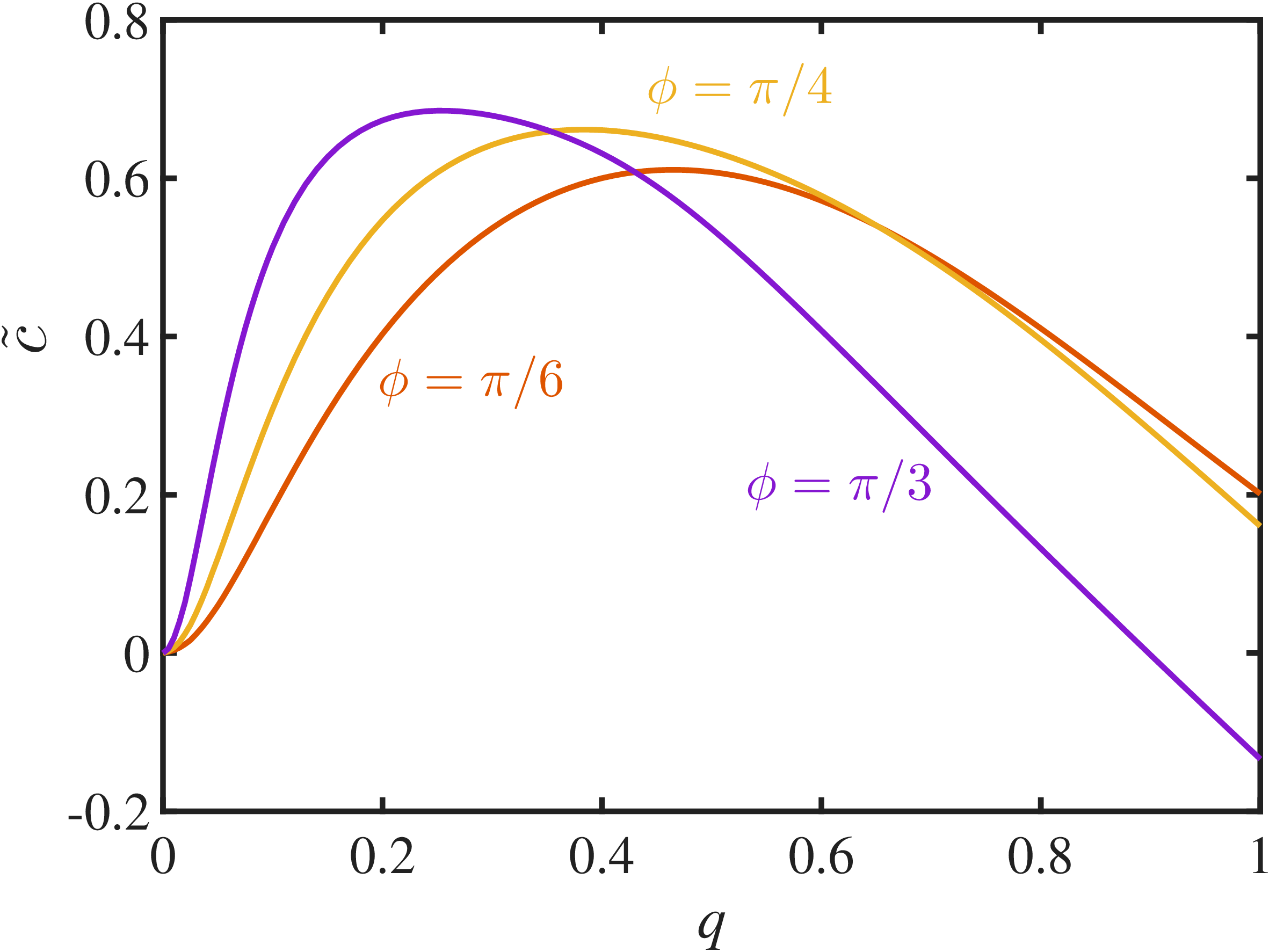}
    \put(-1,70){(b)}
    \end{overpic}    
    \caption{Variation of the traveling velocity (a) $c$ in the present coordinate system and (b) $\tilde{c}=c-(1+\alpha)\tan\phi$ in a coordinate system moving vertically and attached to the contact line. Results are shown as a function of the wavenumber $q$ for representative contact-line inclination angles, with $\alpha=10^{-2}$ and $\theta_e=1$.}
    \label{fig:phase_velocity}
\end{figure}

Different from the standing-wave modes for $\phi=0$, the contact-line inclination gives rise to eigenmodes that can be regarded as traveling waves along the ${y}$ axis. The traveling velocity of the normal mode is defined as $c\equiv-\mbox{Im}(\beta)/q$. Fig.~\ref{fig:phase_velocity}(a) shows the dependence of $c$ on the wavenumber $q$ for different inclination angles. As $q$ increases, one can observe a non-monotonic variation of $c$, which first increases from the long-wave value and then decreases. In the range $0<q<1$, we have $c>0$ and hence the corresponding perturbation waves propagate downward along the contact line in the adopted reference frame. Noting that the reference frame moves normally to the contact line, it is also of interest to know how the perturbation waves behave in a frame attached to the contact line but moving vertically. In this new frame of reference, the old frame translates upward along the contact line with a speed equal to $(1+\alpha)\tan\phi$; therefore, the waves propagate at the velocity $\tilde{c}=c-(1+\alpha)\tan\phi$. As shown in Fig.~\ref{fig:phase_velocity}(b), $\tilde{c}$ vanishes as $q\to0$, indicating that long-wave perturbations remain approximately stationary in the vertically moving frame. For waves with sufficiently short wavelengths, $\tilde{c}$ becomes negative and the perturbations propagate upward.

\subsection{Long-wave asymptotic analysis}\label{subsec:Long wave asymptotic analysis}
In the long-wavelength limit, asymptotic analysis can be performed to provide some analytical results that are useful for interpreting the long-wave instability. In this situation, it can be shown that $G$ and $\beta$ admit the expansions
\begin{gather}
      G({x})=g_{0}({x})+qg_{1}({x})+q^{2}g_{2}({x})+O(q^3), \label{analytical:G}\\
      \beta=q\beta_{1}+q^{2}\beta_{2}+O(q^3), \label{analytical:beta}
\end{gather}
for $q\ll1$. The solutions can be derived by substituting the expansions into Eqs.~\eqref{Linearized_perturbation_eq}--\eqref{Linearized_perturbation_cl_boundary_complete} and collecting terms of the same order in $q$. 

At the leading order, $O(q^0)$, we obtain
\begin{equation}
      \frac{d }{d {x} }\left[h_{0}(h_{0}^{2}+\alpha)g_{0{x} {x} {x} }+2h_{0}^2(h_{0{x}{x}{x}}-1)g_{0} \right]=0.
      \label{complete:zero_order_eq}
\end{equation}
The corresponding boundary conditions are described as
\begin{equation}
      g_0 \to 0,\quad g_{0{x}}\to 0\quad \mathrm{as}\quad {x}\to\infty,
      \label{zero_order_bc_infty}
\end{equation}
and
\begin{equation}
    g_{0} - g_{0{x}}\frac{h_{0{x}}}{h_{0{x}{x}}} = 0,\quad g_{0{x}{x}{x}}=0 \quad\mathrm{at}\quad {x}=0,
    \label{complete:zero_order_bc_cl}
\end{equation}
for both complete wetting and partial wetting. Consistent with \citet{troianFingeringInstabilitiesDriven1989}, the solution of the leading-order eigenfunction is given by
\begin{equation}
    g_0=h_{0{x}},
    \label{eq:g0}
\end{equation}
where the coefficient has been normalized to unity without loss of generality. The corresponding perturbation simply represents a translation of the base film profile in the direction perpendicular to the contact line.

For the first-order approximation, $O(q)$, the governing equation is
\begin{equation}
      \beta_1g_0+{i}g_0(3h_0^2+\alpha)\tan\phi+\frac{d }{d {x} }\left[h_{0}(h_{0}^{2}+\alpha)g_{1{x} {x} {x} }+2h_{0}^2(h_{0{x}{x}{x}}-1) g_{1} \right]=0.
      \label{complete:first_order_eq}
\end{equation}
The corresponding boundary conditions in the far field are
\begin{equation}
      g_1 \to 0,\quad g_{1{x}}\to 0 \quad \mbox{as}\quad {x}\to\infty.
      \label{first_order_bc_2}
\end{equation}
At the contact line (${x}=0$), they are expressed as
\begin{equation}
g_{1} - g_{1{x}}\displaystyle\frac{h_{0{x}}}{h_{0{x}{x}}} = 0,\quad
\end{equation}
and
\begin{empheq}[left={0=\empheqlbrace}]{align}
    & \beta_1 g_0 + \alpha\left( h_{0{x}} g_{1{x}{x}{x}} + {i} g_0 \tan \phi \right)  &&\text{for partial wetting}, \label{eq:g1partial} \\[6pt]
    & \beta_1 g_{0{x}} + \alpha\left( h_{0{x}{x}} g_{1{x}{x}{x}} + {i} g_{0{x}} \tan \phi \right) && \text{for complete wetting}. \label{eq:g1complete}
\end{empheq}
According to Eq.~\eqref{eq:g0}, the boundary conditions given by Eqs.~\eqref{eq:g1partial} and \eqref{eq:g1complete} can be written in a unified form
\begin{equation}
    \beta_1 + \alpha\left(g_{1{x}{x}{x}} + {i} \tan \phi \right)=0 \quad \mbox{at}\quad {x}=0.
\end{equation}
Substituting Eq.~\eqref{eq:g0} into Eq.~\eqref{complete:first_order_eq}, it follows that
\begin{equation}
      \frac{d}{d {x}}\left[(\beta_1+{i}\alpha\tan\phi)h_0+{i}h_0^3\tan\phi+h_0(h_{0}^{2}+\alpha)g_{1{x} {x} {x} }+2h_{0}^2(h_{0{x}{x}{x}}-1)g_{1}  \right] =0.
      \label{complete:first_derivation_1}
\end{equation}
Integrating Eq.~\eqref{complete:first_derivation_1} from ${x}=0$ to $\infty$ and simplifying the resulting expression with boundary conditions of $h_0$ and $g_1$, we obtain
\begin{equation}
      \beta_1=-{i}(1+\alpha)\tan\phi.
      \label{eq:beta_1}
\end{equation}
As a result, the traveling velocity of long waves is given by $c=(1+\alpha)\tan\phi$, or $\tilde{c}=0$, consistent with the numerical results in Fig.~\ref{fig:phase_velocity}. Inspired by the form of $\beta_1$, we assume the eigenfunction $g_1=-{i}f({x})\tan\phi $. Substituting this expression into Eq.~\eqref{complete:first_derivation_1} yields the governing equation for $f({x})$:
\begin{equation}
      \frac{d}{d {x}}\left[h_0(h_0^2+\alpha)(f_{{x}{x}{x}}-h_{0{x}{x}{x}})+2h_0^2(h_{0{x}{x}{x}}-1)f\right]=0.
      \label{first_derivation_2}
\end{equation}
The corresponding boundary conditions are
\begin{gather}
      f\to 0,\quad f_{x}\to 0\quad \text{as}\quad {x}\to \infty.
      \label{first_order_bc_4}
\end{gather}
and
\begin{equation}
    f - f_{{x}}\frac{h_{0{x}}}{h_{0{x}{x}}} = 0,\quad f_{{x}{x}{x}}=-\frac{1}{\alpha} \quad \text{at}\quad {x}=0,
    \label{first_order_bc_3}
\end{equation}
for both complete wetting and partial wetting. In the complete wetting case, the solution is found to be $f={x} h_{0{x}}/3$, leading to 
\begin{equation}
    g_1=-\frac{{i}}{3}{x} h_{0{x}}\tan\phi \quad \mbox{when}\quad \theta_e=0.
\end{equation} 
Conversely, in the case of partial wetting ($\theta_e\neq0$), $f$ and $g_1$ should be determined numerically together with $h_0$.

To obtain the growth rates, it is necessary to further consider the $O(q^2)$ problem. The governing equation of $g_2$ reads
\begin{align}
      \beta_2g_0&+\beta_1g_1-h_0(h_0^2+\alpha)g_{0{x}{x}}+{i}g_1(3h_0^2+\alpha)\tan\phi \notag \\
      &+\frac{d }{d {x} }\left[h_{0}(h_{0}^{2}+\alpha)(g_{2{x} {x} {x} }-g_{0{x}})+2h_0^2(h_{0{x}{x}{x}}-1)g_2 \right]=0,
      \label{second_order_eq}
\end{align}
supplemented by the far-field conditions
\begin{equation}
      g_{2}\to 0,\quad g_{2{x}}\to0 \quad\text{as}\quad {x}\to\infty.
      \label{second_order_bc_2}
\end{equation}
For the present purposes, the boundary conditions at ${x}=0$ are not needed. Integrating Eq.~\eqref{second_order_eq} over the entire domain and using Eqs.~\eqref{base_flow_cl_boundary} and \eqref{second_order_bc_2}, we arrive at
\begin{equation}
      \int_0^{\infty}\left[ \beta_2g_0+\beta_1g_1-h_0(h_0^2+\alpha)g_{0{x}{x}}+{i}g_1(3h_0^2+\alpha)\tan\phi \right] d{x}=0.
      \label{second_derivation_1}
\end{equation}
Substituting the expressions of $g_0$, $g_1$ and $\beta_1$ and adopting Eqs.~\eqref{base_flow_cl_boundary}, \eqref{base_flow_far_boundary} and \eqref{base_flow_eq}, we obtain
\begin{equation}
      \beta_2 = \int_0^{\infty} h_0(h_0^2 - 1)  d{x} + \tan^2\phi \int_0^{\infty} (1-3h_0^2)f  d{x}.
      \label{partial:beta_2}
\end{equation}
In the complete wetting case, where $f({x})$ possesses an explicit expression, the last term can be integrated by parts, giving rise to
\begin{equation}
      \beta_2=\left(1+\frac{1}{3}\tan^2\phi\right)\int_0^{\infty} h_0(h_0^2-1)  d{x}.
      \label{complete:beta_2}
\end{equation}
Since $\beta_2$ is real, the leading order of the growth rate is given by $\beta_2q^2$. For the case of horizontal contact lines, Eqs.~\eqref{partial:beta_2} and \eqref{complete:beta_2} reduce to the result proposed by \citet{troianFingeringInstabilitiesDriven1989}. It can be confirmed that the first integral in Eq.~\eqref{partial:beta_2} is positive owing to the presence of the capillary ridge of the unperturbed profile, more specifically the region $h_0>1$, indicating that long waves at $\phi=0$ are inherently unstable.

\begin{figure}
    \centerline{\includegraphics[width=0.5\textwidth]{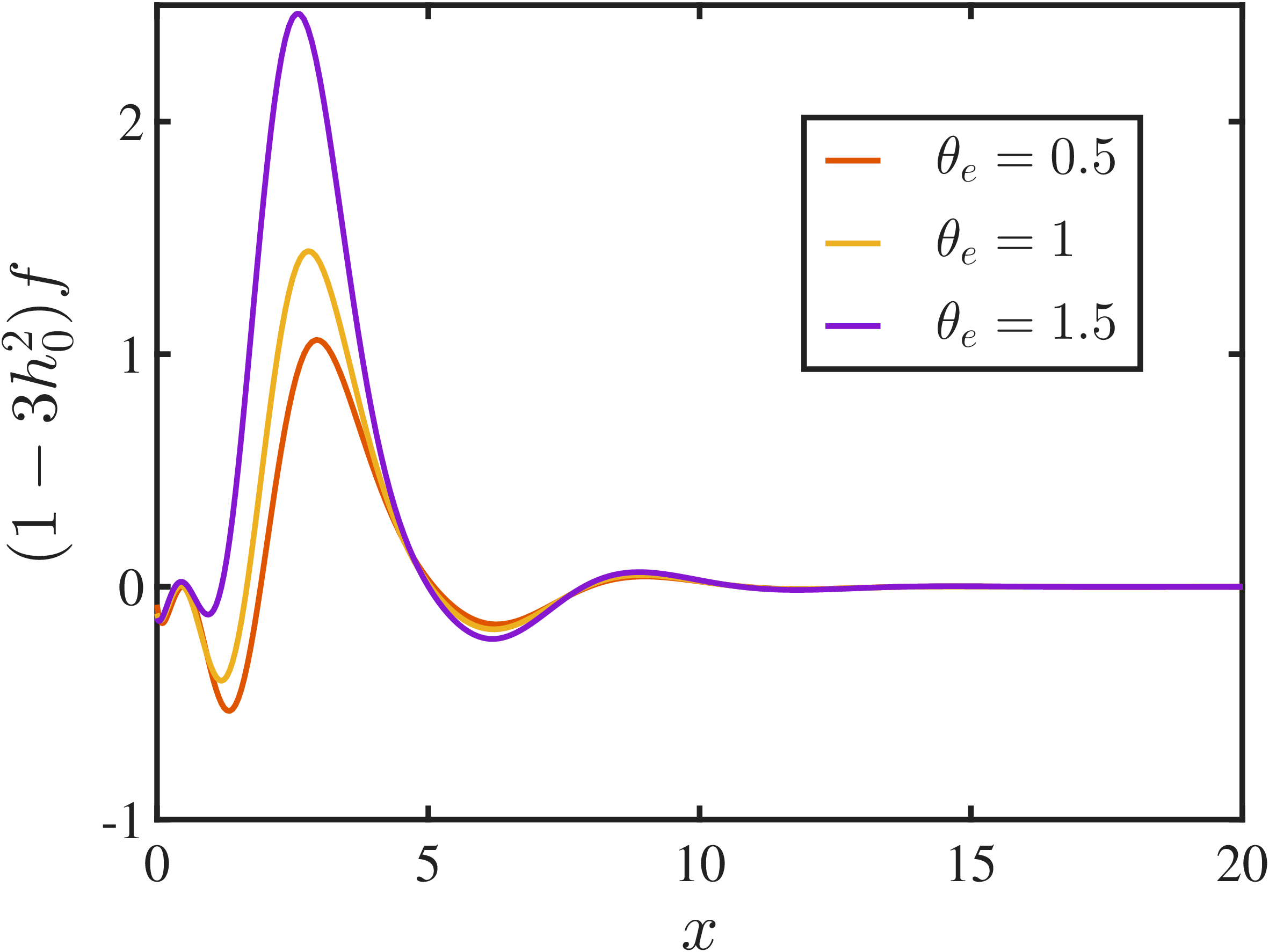}}
    \caption{Distributions of $(1-3h_0^2)f$ versus ${x}$ for representative contact angles $\theta_e$ at $\alpha=0.01$.}
    \label{fig:f_related}
\end{figure}

For a film with an inclined contact line, the growth rates of long-wave perturbations are altered by the terms proportional to $\tan^2\phi$, noting that the integrals in Eqs.~\eqref{partial:beta_2} and \eqref{complete:beta_2} are independent of $\phi$. When $\theta_e=0$, Eq.~\eqref{complete:beta_2} clearly indicates that the growth of long waves is enhanced by the contact-line inclination, since $\beta_2$ increases with $\phi$. For the case of partial wetting ($\theta_e>0$), the second integral in Eq.~\eqref{partial:beta_2} can also be shown to be positive, such that the inclination likewise promotes the long-wave instability. We show in Fig.~\ref{fig:f_related} the integrand of the second integral for representative values of $\theta_e$. It is seen that the primary contribution to the integration comes from the region within the capillary ridge shown in Fig.~\ref{fig:base_flow_profile}.

\begin{figure}
  \centering
  \begin{overpic}[width=0.45\textwidth]{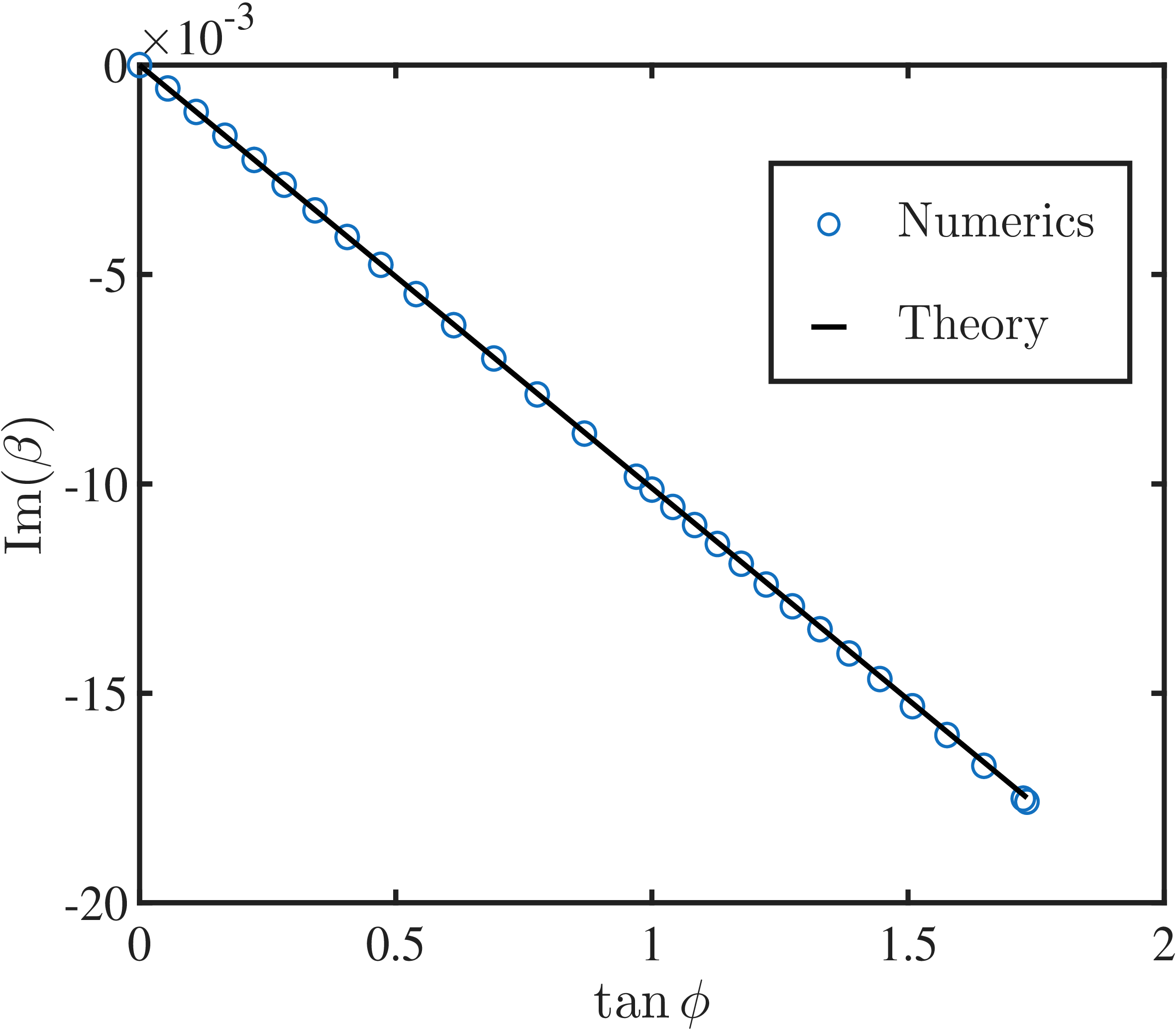}
    \put(-5,70){(a)}
  \end{overpic}
  \quad
  \begin{overpic}[width=0.45\textwidth]{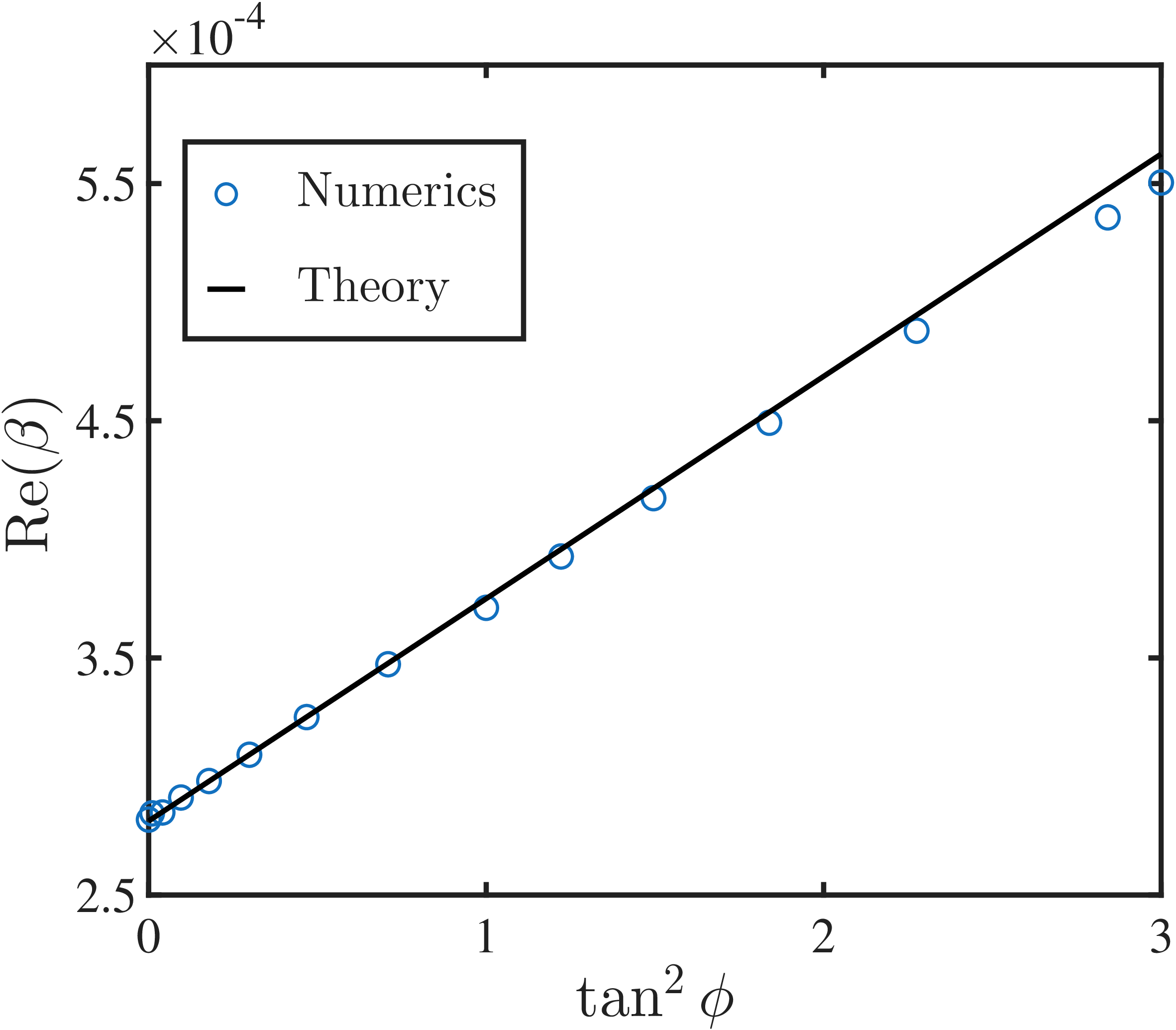}
    \put(-1,70){(b)}
  \end{overpic}
  \begin{overpic}[width=0.45\textwidth]{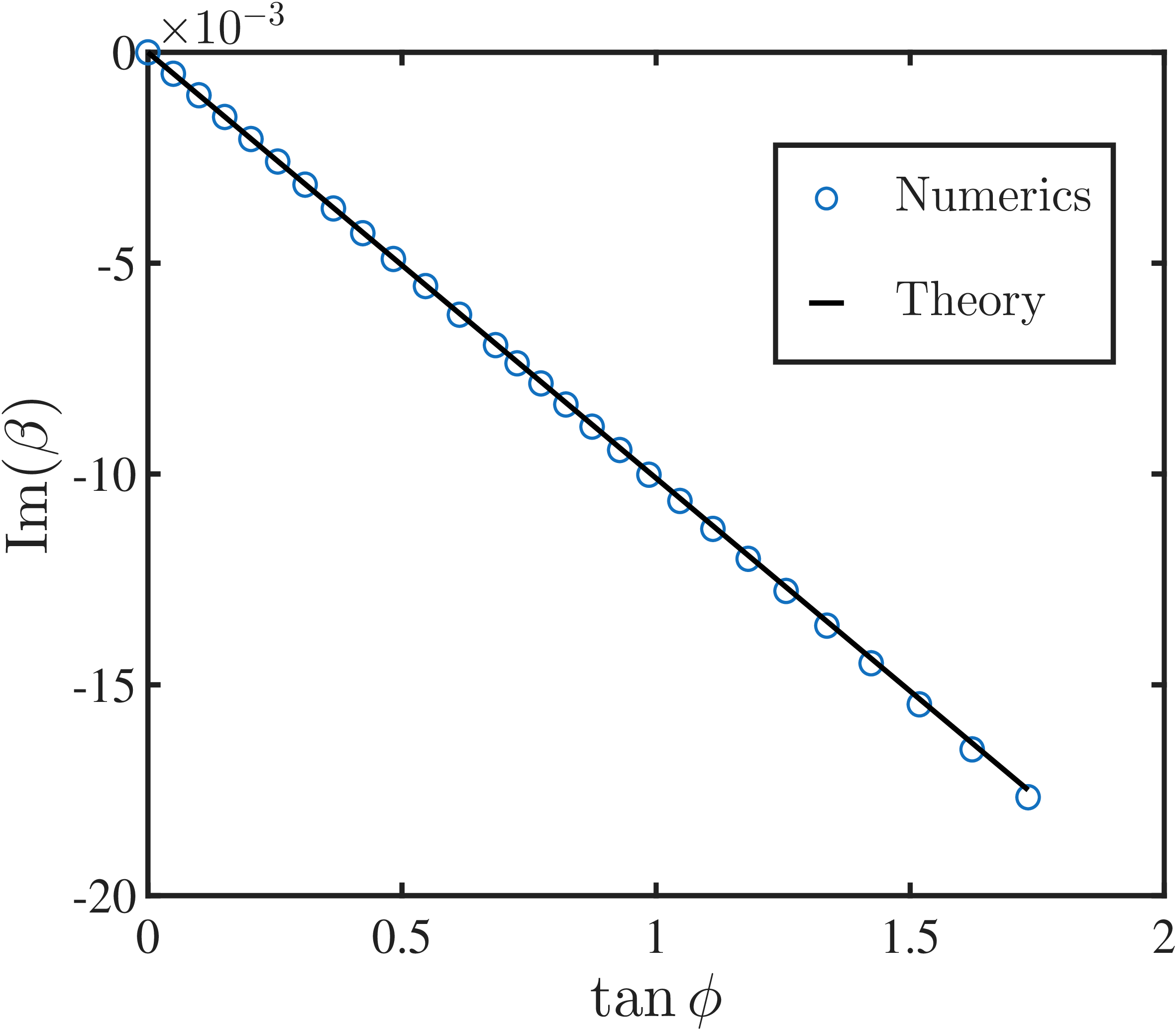}
    \put(-1,70){(c)}
  \end{overpic}
  \quad
  \begin{overpic}[width=0.45\textwidth]{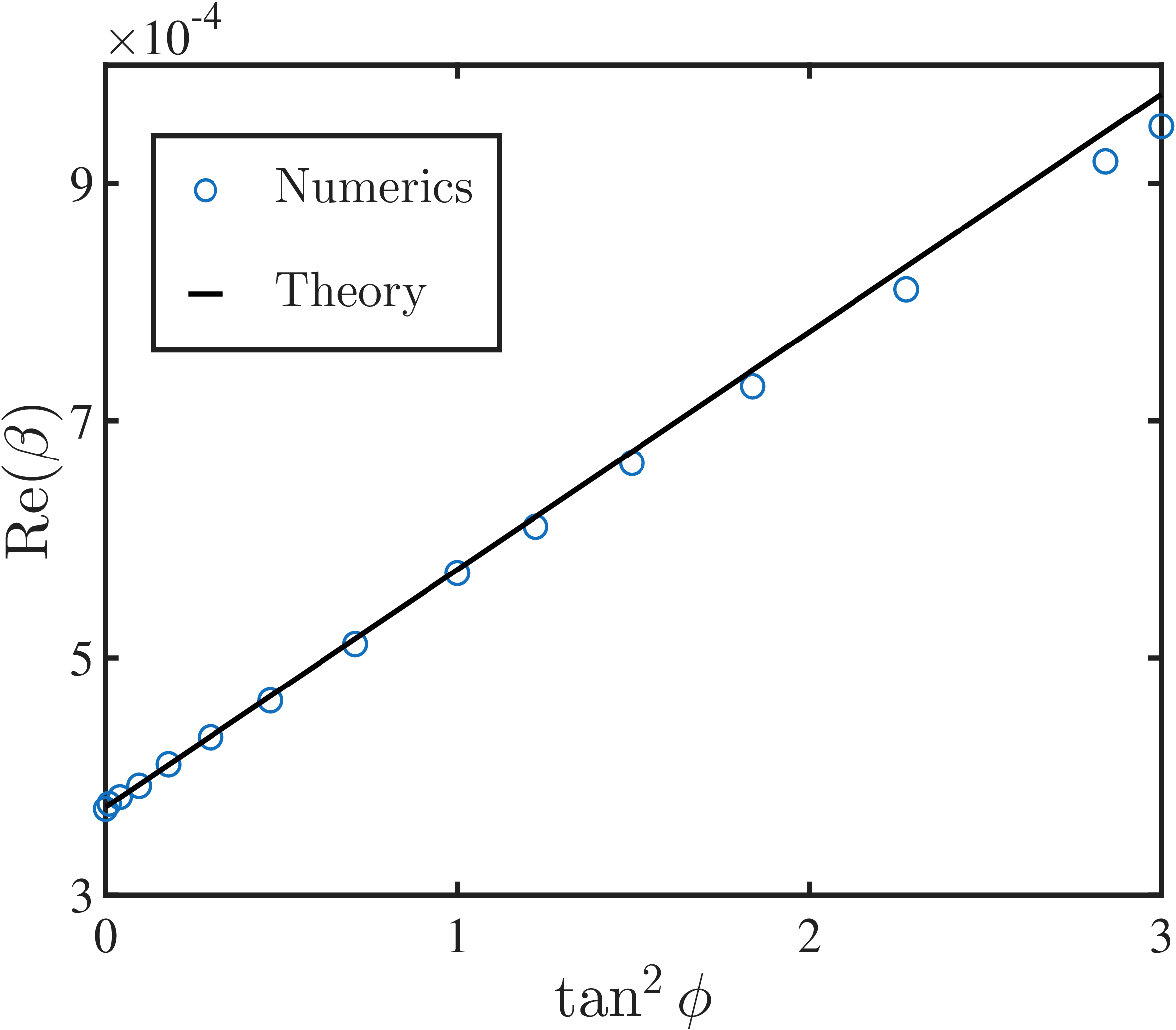}
    \put(-1,70){(d)}
  \end{overpic}
  \caption{Comparison between numerical results and long-wave theory of $\beta$ at $\alpha=0.01$ and $q=0.01$. (a, b) $\theta_e=0$; (c, d) $\theta_e=1$. The solid lines in (a, c) correspond to Eq.~\eqref{eq:beta_1}; the solid lines in (b) and (d) correspond to Eqs.~\eqref{partial:beta_2} and \eqref{complete:beta_2}, respectively.}
  \label{result:long_wave_analysis_complete}
\end{figure}

The asymptotic theory of long-wave instability is validated against the numerical results of the eigenvalue problem in Fig.~\ref{result:long_wave_analysis_complete}, which illustrates the relation between $\beta$ and $\phi$. The contact angle is selected as $\theta_e=0$ and 1, representing the case of complete and partial wetting, respectively. The wavenumber is $q=0.01$, a value sufficiently small to satisfy the long-wave approximation. In agreement with the asymptotic theory, the imaginary part of $\beta$ scales with $\tan\phi$, while its real part varies linearly with $\tan^2\phi$. The numerical results of the eigenvalue problem demonstrate excellent consistency with the theoretical predictions, confirming the validity of the analytical model under both wetting regimes.

\section{\label{sec:Mechanisms} Mechanisms}
The underlying mechanisms of the conventional fingering instability with a horizontal contact line have been investigated from the viewpoints of disturbance velocity and energy balance \citep{brennerinstabilitymechanismdriven1993,spaidStabilityNewtonianViscoelastic1996}. Here, we propose an interpretation of the influence of the contact-line inclination, specifically the destabilizing effect in the long-wavelength region and the stabilizing effect in the finite-wavelength region, based on the numerical and asymptotic results.
\begin{figure}
 \centering
 \begin{overpic}[width=0.55\textwidth]{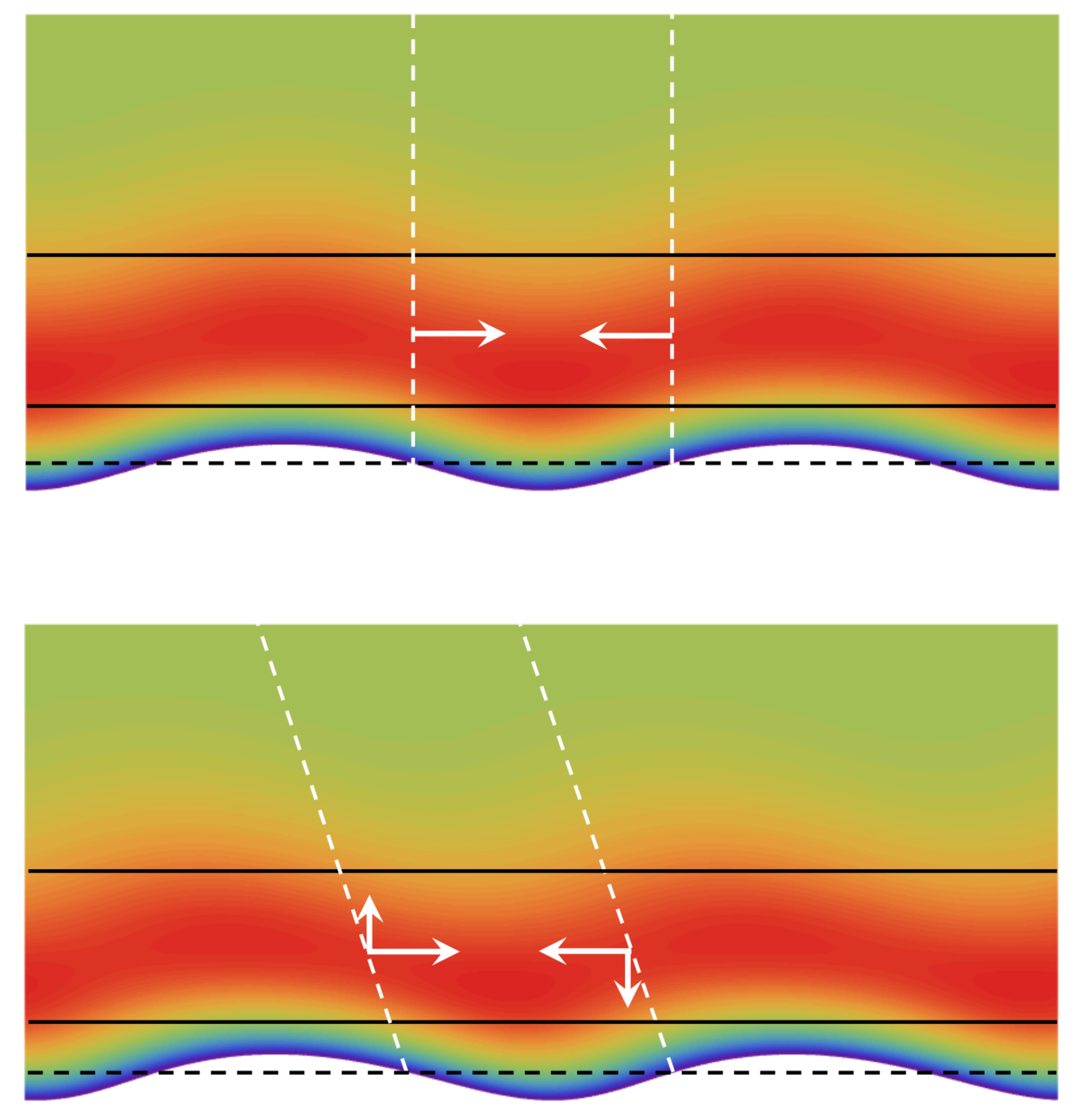}
    \put(-8,95){(a)}
    \put(-8,40){(b)}
  \end{overpic}
  \caption{Sketch for the long-wave instability mechanism when the contact line is (a) horizontal and (b) inclined. The contours represent the perturbed film thickness, with the red region denoting the capillary ridge. The black dashed and solid lines represent the positions of $h_0=0$ (the contact line) and $h_0=1$ for the basic profile, respectively. The white dashed lines represent the boundaries between the crest and trough of the perturbations. For long waves, the local translation of the basic film profile yields a variation of the curvature and hence a pressure disturbance. In the region bounded by the solid lines ($h_0>1$), this pressure disturbance drives a flow from the crests to the troughs of the capillary ridge as indicated by the arrows, giving rise to the fingering instability. The contact-line inclination enhances the instability by introducing additional fluxes.}
  \label{fig:mechanism_long-wave}
\end{figure}

\subsection{Long-wavelength region}

\begin{figure}
 \centering
 \begin{overpic}[width=0.55\textwidth]{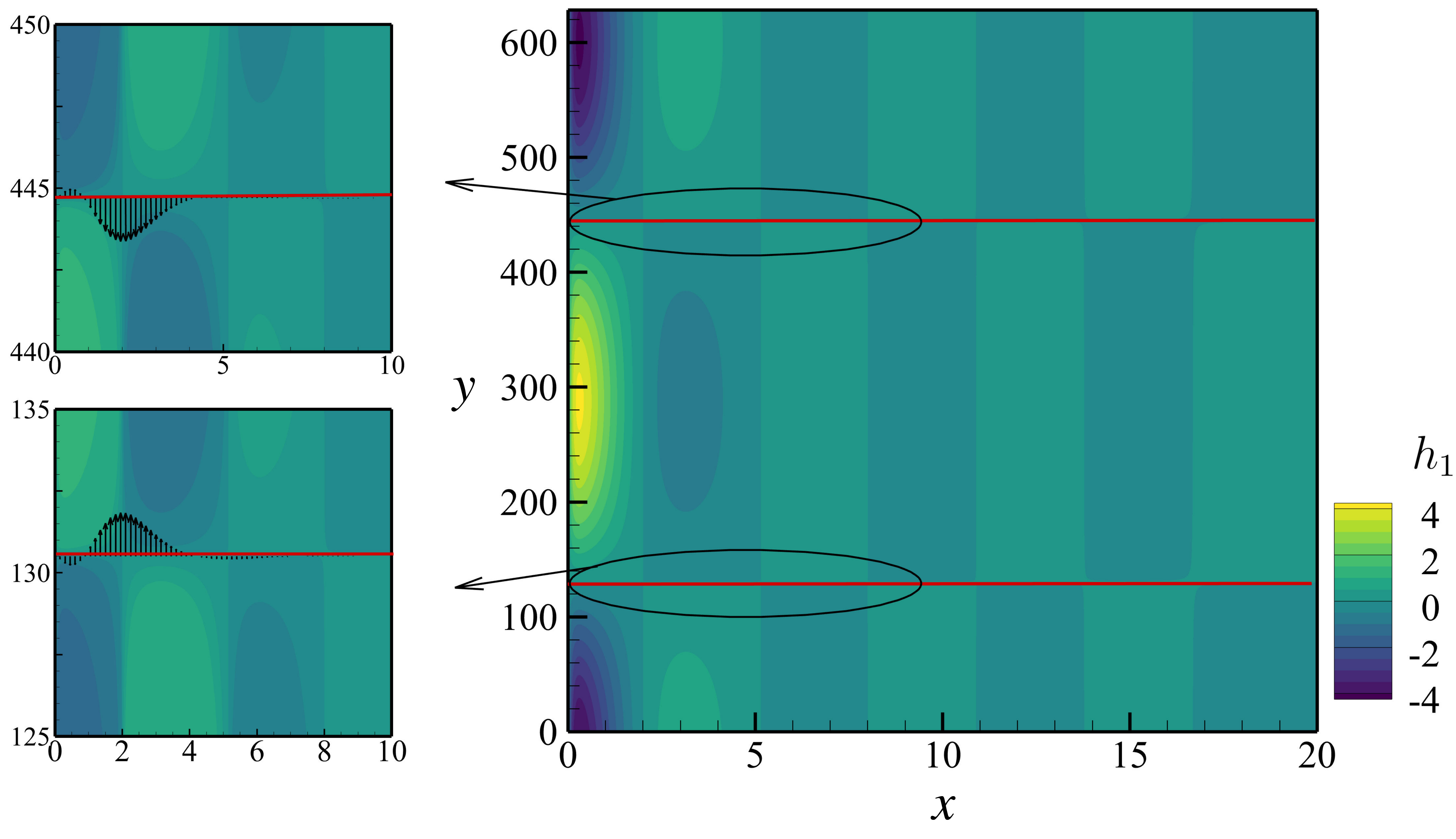}
    \put(-8,50){(a)}
  \end{overpic}

  \begin{overpic}[width=0.55\textwidth]{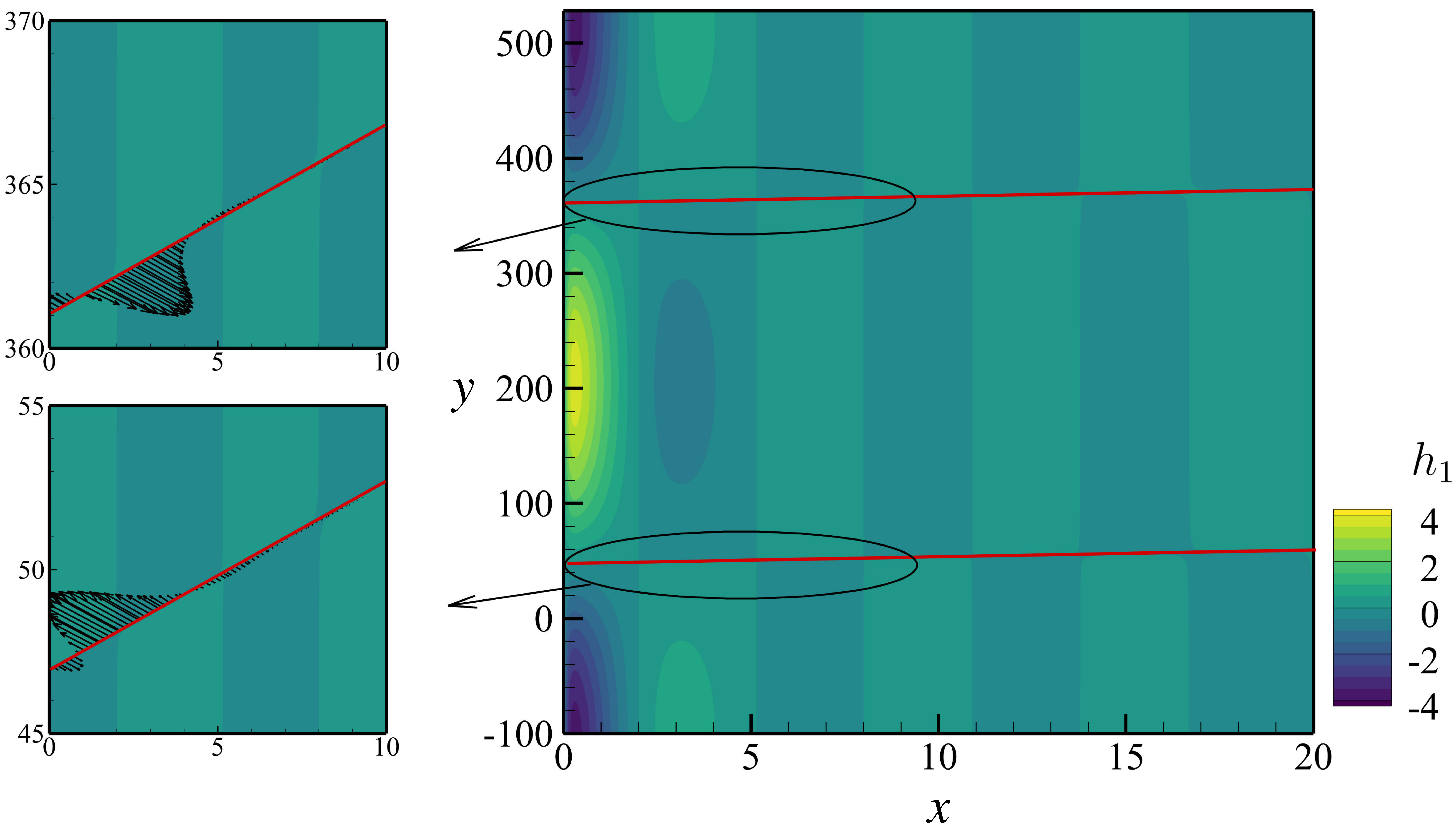}
    \put(-8,50){(b)}
  \end{overpic}
  \caption{Distribution of $h_1$ for (a) $\phi=0$ and (b) $\phi=\pi/3$, with $\alpha=0.01$, $\theta_e=0$ and $q=0.01$. The arrows represent the local flux of velocity perturbation $\boldsymbol{Q}'$ at the red solid lines, which divide the crests and troughs and enclose a semi-period of the wave pattern. The insets magnify the regions near the capillary ridge with an aspect ratio of 1.}
  \label{fig:flux-analysis_long-wave}
\end{figure}

An intuitive interpretation for the long-wave instability is provided in Fig.~\ref{fig:mechanism_long-wave}, where we focus on the region of the capillary ridge, which is mostly responsible for the instability. The growth or decay of the film thickness can be rationalized by analyzing the flux induced by the capillary pressure of the perturbations.

We first revisit the case of a horizontal contact line, as shown in Fig.~\ref{fig:mechanism_long-wave}(a). The perturbed film is characterized by a sinuous capillary ridge. According to the asymptotic analysis, the long-wave perturbation is simply associated with a local translation of the base film profile, either upward or downward. Since the curvature of the base profile, $h_{0{x}{x}}$, is not uniform, this translation would give rise to a curvature perturbation, ${x}_ch_{0{x}{x}{x}}$, where the contribution of the spanwise variations to the curvature has been neglected for long waves. Correspondingly, the pressure perturbation is given by $-{x}_ch_{0{x}{x}{x}}$. The sign of the pressure disturbance is thus determined by $(1-h_0){x}_c$ according to Eq.~\eqref{base_flow_eq}. In the region of $h_0>1$ (bounded by the two solid lines), the capillary pressure thus increases in the crest (${x}_c>0$) and decreases in the trough (${x}_c<0$). The resulting pressure gradient drives a transverse flow from the crest to the trough. While the flow direction may be reversed elsewhere, the net flux is directed to the troughs in the dominant region of $h_0>1$. Consequently, the troughs become more massive and move faster under gravity, leading to the growth of the perturbations. We note that this scenario of capillary ridge instability is different from the well-known Rayleigh--Plateau instability of a cylindrical liquid jet \citep{plateauExperimentalTheoreticalResearches1857,rayleighInstabilityJets1878}. In the latter, the curvature of the base state is uniform and the driving force comes from the curvature gradient induced by varicose rather than sinuous perturbations. In contrast, the non-uniform curvature of the capillary ridge plays a crucial role, enabling the occurrence of sinuous eigenmodes for the fingering instability.

In the presence of an inclined contact line, the long-wave perturbations can also be regarded as a local translation of the base film profile. However, the sinuous capillary ridge tends to be further distorted by the transverse component of gravity, which points to the left in Fig.~\ref{fig:mechanism_long-wave}(b), and the resistance effect of the contact line. The perturbation waves are expected to be sheared such that the boundaries between crests and troughs, represented by the white dashed lines, tilt to the left. As a result, in addition to the transverse flow, the pressure difference can drive a streamwise flow from the crests to the troughs (see Fig.~\ref{fig:mechanism_long-wave}b), promoting the growth of long-wave perturbations. 

The instability mechanism proposed above is confirmed by the numerical results of the eigenvalue problem. Fig.~\ref{fig:flux-analysis_long-wave} presents the distribution of the perturbation $h_1$ with $q=0.01$, which is sufficiently small to represent the long-wavelength region. Since variations in the contact angle do not qualitatively affect the nature of the linear instability, we choose $\theta_e=0$ as a representative case. The inclination angles of the contact line are selected as $\phi=0$ and $\pi/3$, corresponding to the configuration of a horizontal and inclined contact line, respectively. In either case, the eigenmode exhibits a two-dimensional pattern of waves, with the crests and troughs occurring alternately in both streamwise and spanwise directions. The borders between the crests and the troughs can be identified as the red straight lines, which are perpendicular to the contact line when $\phi=0$ and become tilted with respect to it when $\phi=\pi/3$, as shown in the insets. The arrows represent the flux of the velocity perturbations $\boldsymbol{Q}'$ along these borders. Consistent with the qualitative analysis in Fig.~\ref{fig:mechanism_long-wave}, it can be observed that the disturbance flow is predominantly distributed in the capillary ridge region. This flow drives the liquid to the regions around the peaks of $h_1$, which is associated with the protrusion of the contact line shown in Fig.~\ref{fig:mechanism_long-wave}, giving rise to the instability. The main contribution to the net flux originates from the region of $h_0>1$. Moreover, as indicated in Fig.~\ref{fig:flux-analysis_long-wave}(b), the net flow is evidently enhanced for the configuration with an inclined contact line compared to the horizontal contact line, indicating the stronger instability with contact-line inclination.

These results are further rationalized by the long-wave asymptotic solutions presented in Sec.~\ref{subsec:Long wave asymptotic analysis}. We can retrieve the growth rate of the eigenmode by analyzing the mass conservation in the control volume bounded by the two red lines in Fig.~\ref{fig:flux-analysis_long-wave}. For $\theta_e=0$, the form of $h_1$ can be determined by $O(1)$ and $O(q)$ asymptotic solutions, and reads
\begin{equation}
    h_1=\mathrm{e}^{q^2\beta_2\tau}h_{0{x}}\cos\left[q\left({y}-\frac{1}{3}{x}\tan\phi-(1+\alpha)\tau\tan\phi\right)\right]+O(q^2).
    \label{eq:flux_h1}
\end{equation}
Accordingly, the red lines in Fig.~\ref{fig:flux-analysis_long-wave} are simply the contour of phase along which $h_1=0$; they are indeed straight with a slope of $(\tan\phi)/3$. Recalling Eq.~\eqref{dimensionless-flux}, the flux perturbation at these borders is given by
\begin{equation}
    \boldsymbol{Q}'=\epsilon h_0(h_0^2+\alpha)\nabla h_{1{x}{x}}+O(\epsilon q^2).
\end{equation}
Evaluating the flux using Eq.~\eqref{eq:flux_h1} and integrating along the two borders, we can obtain the net flux into the control volume as
\begin{equation}
      Q_{\text{net}} = 2\int_0^\infty \boldsymbol{Q}' \cdot \boldsymbol{n}^{\ast} dl
      =2\epsilon q \mathrm{e}^{q^2\beta_2\tau}\left(1+\frac{1}{3}\tan^2\phi \right) \int_0^{\infty} h_0(h_0^2 - 1) d{x} + O(\epsilon q^2),
      \label{eq:flux_Q_net}
\end{equation}
where $\boldsymbol{n}^{\ast}$ is the inward unit normal and $l$ is the arc length measured from the contact line. It can be demonstrated that the contributions from the contact line and the far field to the net flux are negligible. On the other hand, the perturbed film corresponds to a variation of the liquid volume given by
\begin{equation}
    V'=\int_S \epsilon h_1 d{x} d{y}=2\epsilon q^{-1}\beta_2 \mathrm{e}^{q^2\beta_2\tau} + O(\epsilon q^2),
\end{equation}
where $S$ denotes the projected area of the control volume. Then it is straightforward to use the mass conservation condition $dV'/d\tau=Q_{\text{net}}$ to obtain the growth rate $\beta_2$, which is the same as Eq.~\eqref{complete:beta_2}.

\subsection{Finite-wavelength region}

\begin{figure}
 \centering
 \begin{overpic}[width=0.45\textwidth]{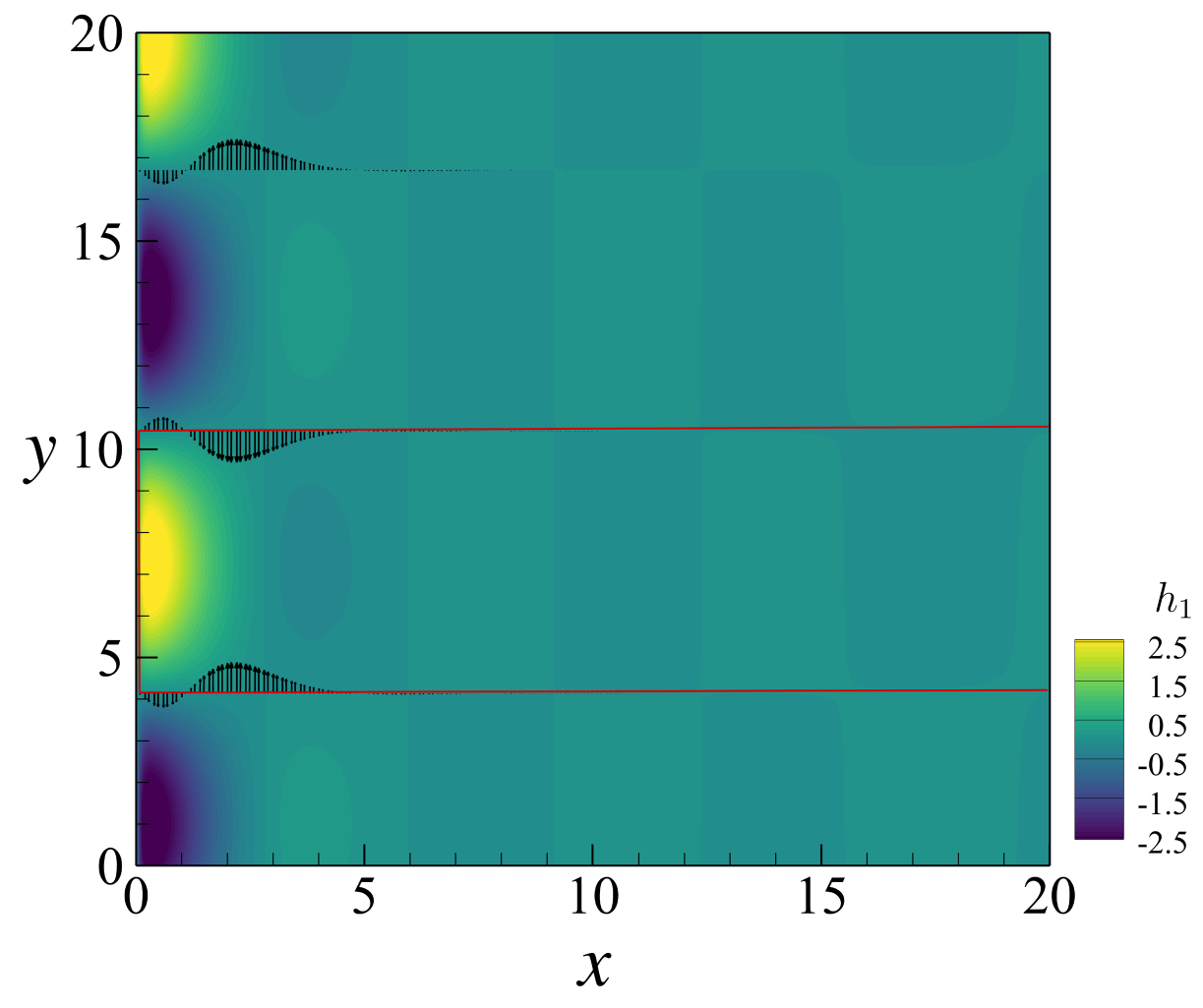}
    \put(-2,78){(a)}
  \end{overpic}
 \begin{overpic}[width=0.47\textwidth]{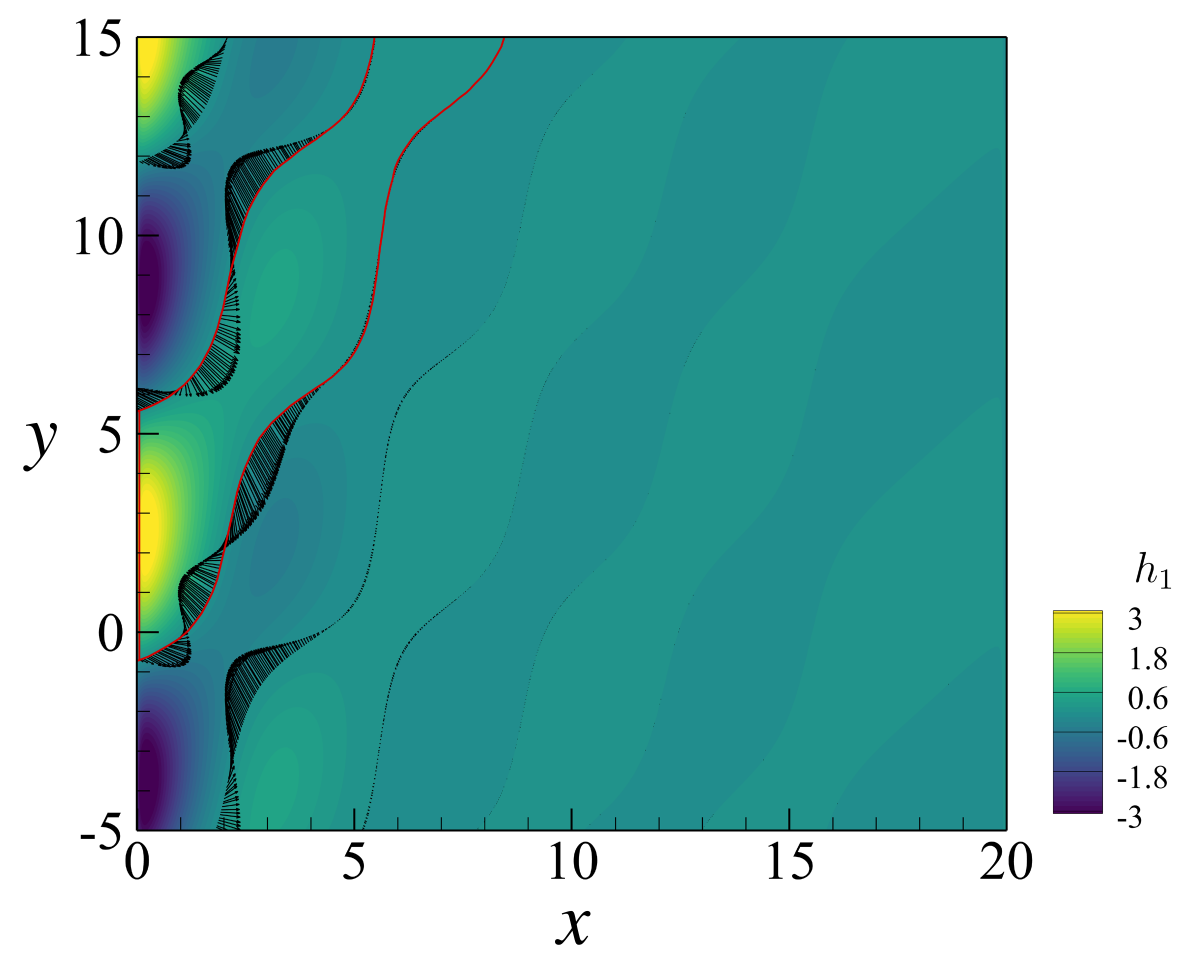}
    \put(-2,75){(b)}
  \end{overpic}
  \caption{Distribution of $h_1$ for (a) $\phi=0$ and (b) $\phi=\pi/4$, with $\alpha=0.01$, $\theta_e=0$ and $q=0.5$. The arrows represent the local flux of velocity perturbation $\boldsymbol{Q}'$ at the red solid lines, which divide the crests and troughs and enclose a semi-period of the wave pattern.}
  \label{fig:flux-analysis_finite-wave}
\end{figure}

In the finite-wavelength region, it is difficult to construct an intuitive interpretation of the stabilizing effect of the contact-line inclination, since analytical solutions are unavailable. Thus we elucidate the mechanism by performing the flux analysis based on numerical solutions.

Fig.~\ref{fig:flux-analysis_finite-wave} demonstrates the distribution of the perturbation $h_1$ and the corresponding flux of perturbation $\boldsymbol{Q}'$ along the equiphase lines ($h_1=0$) for $\theta_e=0$, $q=0.5$ and two representative values of the inclination angle. According to our numerical results of the eigenvalue problem, the eigenmode grows for $\phi=0$ and decays for $\phi=\pi/4$. We can employ a control volume defined in the same manner as in the long-wavelength flux analysis. 
As shown in Fig.~\ref{fig:flux-analysis_finite-wave}(a), for the case of $\phi=0$, one can observe a net inflow flux across the red solid lines, indicating a growing eigenmode. For films with inclined contact lines, the control volume becomes tilted with respect to the contact line and is bounded by borders with irregular shapes. More importantly, the flow perturbation within the capillary ridge changes direction more times along each border. For $\phi=\pi/4$, the outward flow is sufficiently strong, giving rise to an overall outflow flux and hence the decay of the eigenmode.

\section{\label{sec:Conclusions}Conclusions}
In this paper, we examine the stability of a gravity-driven thin film with an inclined contact line, a configuration naturally produced by rotating the plate by a prescribed angle from the classical horizontal-contact-line setup. In the framework of lubrication theory, we formulated a dimensionless evolution equation for the film thickness and performed a linear stability analysis to assess the influence of the contact-line inclination on the instability characteristics. 
Compared to the horizontal contact line configuration, the inclination of the contact line introduced a spanwise flux, which would redefine the flow field of the base state and fundamentally alter the perturbation dynamics.

Regarding the evolution of the perturbations, a notable feature introduced by the inclination is that disturbances propagate as traveling waves along the contact line across the entire wavelength spectrum, in contrast to the stationary wave observed for a horizontal contact line. While this propagation behavior represents a distinctive trait of the inclined system, the modification of perturbation growth rates across different wavelength regions is of particular concern in the present study. Owing to the stabilizing effect of surface tension, short-wavelength perturbations remain continuously damped; consequently, the unstable finite- and long-wavelength regions emerge as the main regimes of physical significance governing the film evolution.

For finite-wavelength perturbations, the linear stability analysis demonstrates that contact-line inclination significantly reduces the growth rate and narrows the unstable bandwidth, thereby stabilizing the flow. A detailed flux analysis reveals that this stabilization arises from a complex redistribution of the perturbation flux within the capillary ridge, which leads to a net outflow from the region of contact line protrusion and thus suppresses the growth. Furthermore, the most unstable wavenumber shifts progressively toward longer wavelengths as the inclination angle increases. This trend may substantially alter the eventual film morphology, resulting in a larger spacing between adjacent finger-like structures at steeper inclinations.

In the long-wave limit, the perturbations exhibit a slight enhancement in growth rate as the inclination angle increases. This enhanced instability can be explicitly described by the asymptotic analysis, which shows that the growth rate for a horizontal contact line is augmented by an additional contribution proportional to $\tan^2\phi$. 
Analysis of the growth-rate expression reveals that the long-wavelength instability is associated with the presence of the capillary ridge in the base-state film profile. From a flux perspective, this instability is attributed to a shearing effect introduced by the inclined contact line, which tilts the boundaries between crests and troughs. This tilt generates an additional streamwise flux that promotes the accumulation of liquid in the regions of protruding contact lines and thereby enhances the instability.

The findings of this study extend the classical understanding of fingering instabilities by demonstrating that contact-line inclination, or a tangential body force, can serve as an effective controlling parameter. Since the physical configuration considered here represents a local approximation of the film region away from the plate boundaries, a potential direction for future work is to investigate edge effects. Furthermore, as the present analysis is limited to linear instability, it would be valuable to explore the subsequent nonlinear evolution of the fingers under inclined contact-line conditions and determine how the linear predictions manifest in the fully developed patterns.

% \begin{bmhead}[Funding.]

% \end{bmhead}

% \begin{bmhead}[Declaration of interests.]
% The authors report no conflict of interest.
% \end{bmhead}
\begin{acknowledgments}
This work was supported by the NSFC with grant Nos. 12325208, 12241204 and 12388101, and China Manned Space Program with grant No. TGMTYY14026.

% The authors report no conflict of interest.
\end{acknowledgments}

\section*{data availability}
The data that support the findings of this article are not publicly available. The data are available from the authors upon reasonable request.

\appendix

\section{\label{boundary_condition}Boundary conditions of the linearized equation at the contact line}

The contact-line position is subjected to a small perturbation
\begin{equation}
      {x}_c = \epsilon \Lambda({y},\tau), \quad \text{with}\quad \epsilon\ll1,
      \label{xi_c}
\end{equation}
and the corresponding unit normal is given by
\begin{equation}
      \boldsymbol{n} = \frac{(-1, \epsilon\Lambda_{{y}})}{(1 + \epsilon^2\Lambda_{y}^2)^{1/2}}.
      \label{normal_vector}
\end{equation}
According to the condition of vanishing film thickness in Eq.~\eqref{contact_line_boundary}, we have
\begin{equation}
      h = h_0 + \epsilon h_{1} = 0\quad\text{at}\quad{x}={x}_c,
      \label{Linearized_perturbation_cl_h}
\end{equation}
which can be lcinearized with respect to ${x}=0$ to give
\begin{equation}
      h_1 = -\Lambda h_{0{x}}\quad\text{at}\quad {x}=0.
      \label{chara_cl_pre_1}
\end{equation}
The contact angle condition in Eq.~\eqref{contact_line_boundary} can also be linearized as
\begin{equation}
      h_{1{x}} = -\Lambda h_{0{x}{x}}\quad\text{at}\quad {x}=0.
      \label{chara_cl_pre_2}
\end{equation}
The final condition to be considered is the kinematic condition Eq.~\eqref{BC_kinematic_condition} at ${x}={x}_c$. Substituting Eqs.~\eqref{dimensionless-flux}, \eqref{perturbation_form}, \eqref{xi_c} and \eqref{normal_vector} into Eq.~\eqref{BC_kinematic_condition}, and neglecting higher-order terms, we obtain
\begin{align}
      \Lambda_{\tau} = \alpha(h_{1{x}{x}{x}} + h_{1{y}{y}{x}} - \Lambda_{{y}} \tan\phi)\quad\text{at}\quad {x}=0.
      \label{chara_cl_pre_3}
\end{align}
Eqs.~\eqref{chara_cl_pre_1}, \eqref{chara_cl_pre_2} and \eqref{chara_cl_pre_3} constitute the linearized conditions at the contact line. 

Furthermore, we consider a normal-mode form of $\Lambda({y},\tau)$, i.e.,
\begin{equation}
      \Lambda({y},\tau) = \mbox{Re}\left\{\Lambda_0 e^{{i}q{y} + \beta\tau}\right\}.
      \label{regular_mode_2}
\end{equation}
Based on the normal-mode forms of $h_1$ and $\Lambda$, Eqs.~\eqref{chara_cl_pre_1}, \eqref{chara_cl_pre_2} and \eqref{chara_cl_pre_3} can be reformulated as
\begin{equation}
    G=-\Lambda_0h_{0{x}},
    \label{eq:G}
\end{equation}
\begin{equation}
    G_{x}=-\Lambda_0h_{0{x}{x}},
    \label{eq:Gxi}
\end{equation}
\begin{equation}
    \beta\Lambda_0=\alpha(G_{{x}{x}{x}}-q^2G_{x}-{i}q\Lambda_0\tan\phi),
    \label{eq:Lambda}
\end{equation}
at ${x}=0$. It can be shown that $\Lambda_0$ can be eliminated by combining Eqs.~\eqref{eq:G} and \eqref{eq:Gxi}, which yields
\begin{equation}
      G - G_{{x}}\frac{h_{0{x}}}{h_{0{x}{x}}} = 0.
      \label{chara_cl_1}
\end{equation}
Similarly, for the case of partial wetting ($h_{0{x}}=\theta_e\neq0$ at ${x}=0$), combining Eqs.~\eqref{eq:G} and \eqref{eq:Lambda} leads to
\begin{equation}
      \beta G + \alpha\left[h_{0{x}} (G_{{x}{x}{x}} - q^2 G_{{x}}) + {i}qG \tan\phi  \right] = 0.
      \label{chara_cl_2}
\end{equation}
For the case of complete wetting ($h_{0{x}}=\theta_e=0$ at ${x}=0$), Eqs.~\eqref{eq:Gxi} and \eqref{eq:Lambda} result in
\begin{align}
      \beta G_{{x}} + \alpha \left[h_{0{x}{x}}(G_{{x}{x}{x}} - q^2 G_{{x}}) + {i}qG_x \tan\phi   \right] = 0.
      \label{chara_cl_2_complete}
\end{align}
Notably, Eq.~\eqref{chara_cl_2} remains valid for the complete wetting case by substituting $\theta_e = 0$.

\section{\label{numerical_method}Numerical method}
\begin{figure}
  \centering
  \begin{overpic}[width=0.45\textwidth]{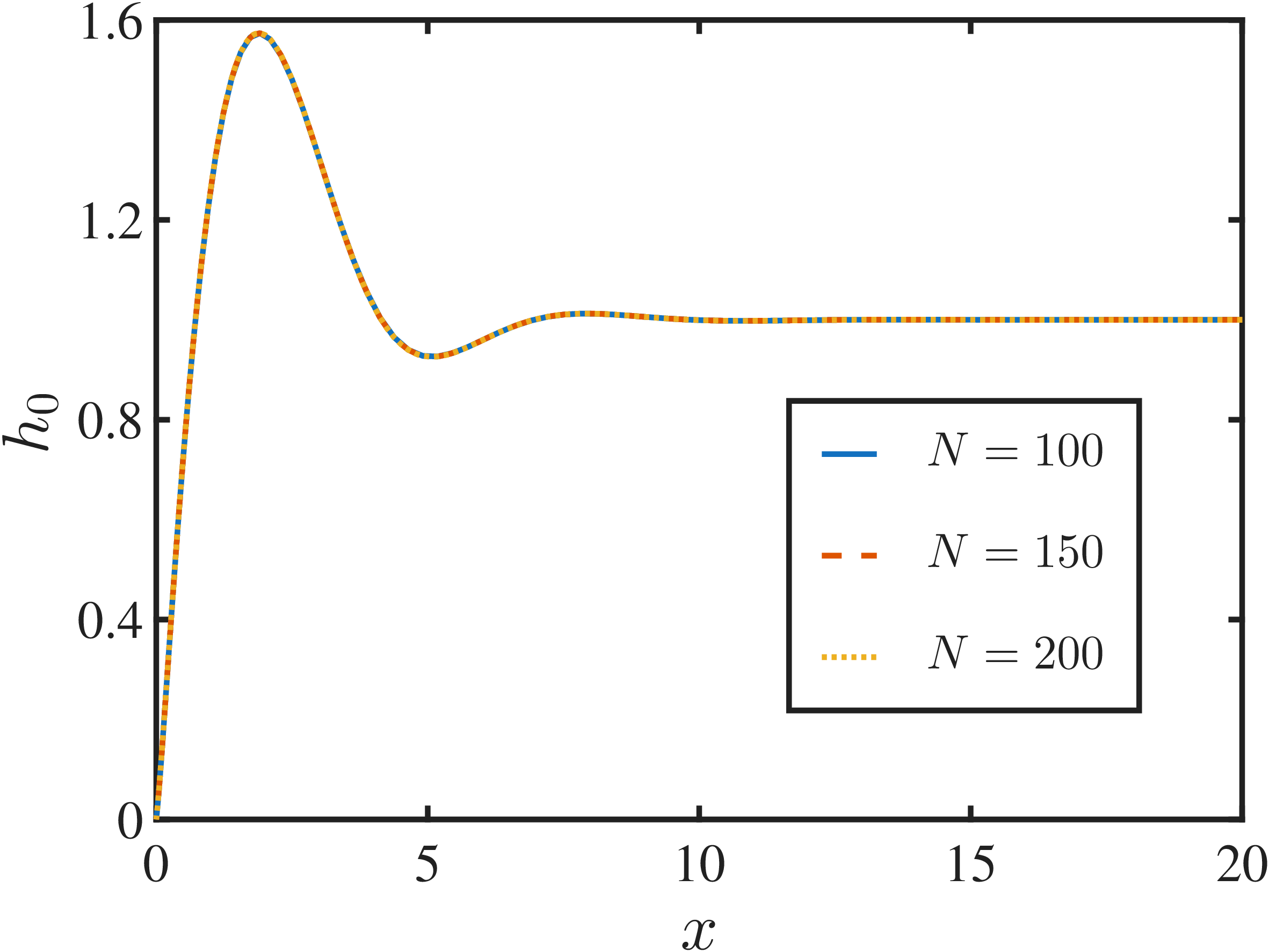}
    \put(-3,70){(a)}
  \end{overpic}
  \quad
  \begin{overpic}[width=0.45\textwidth]{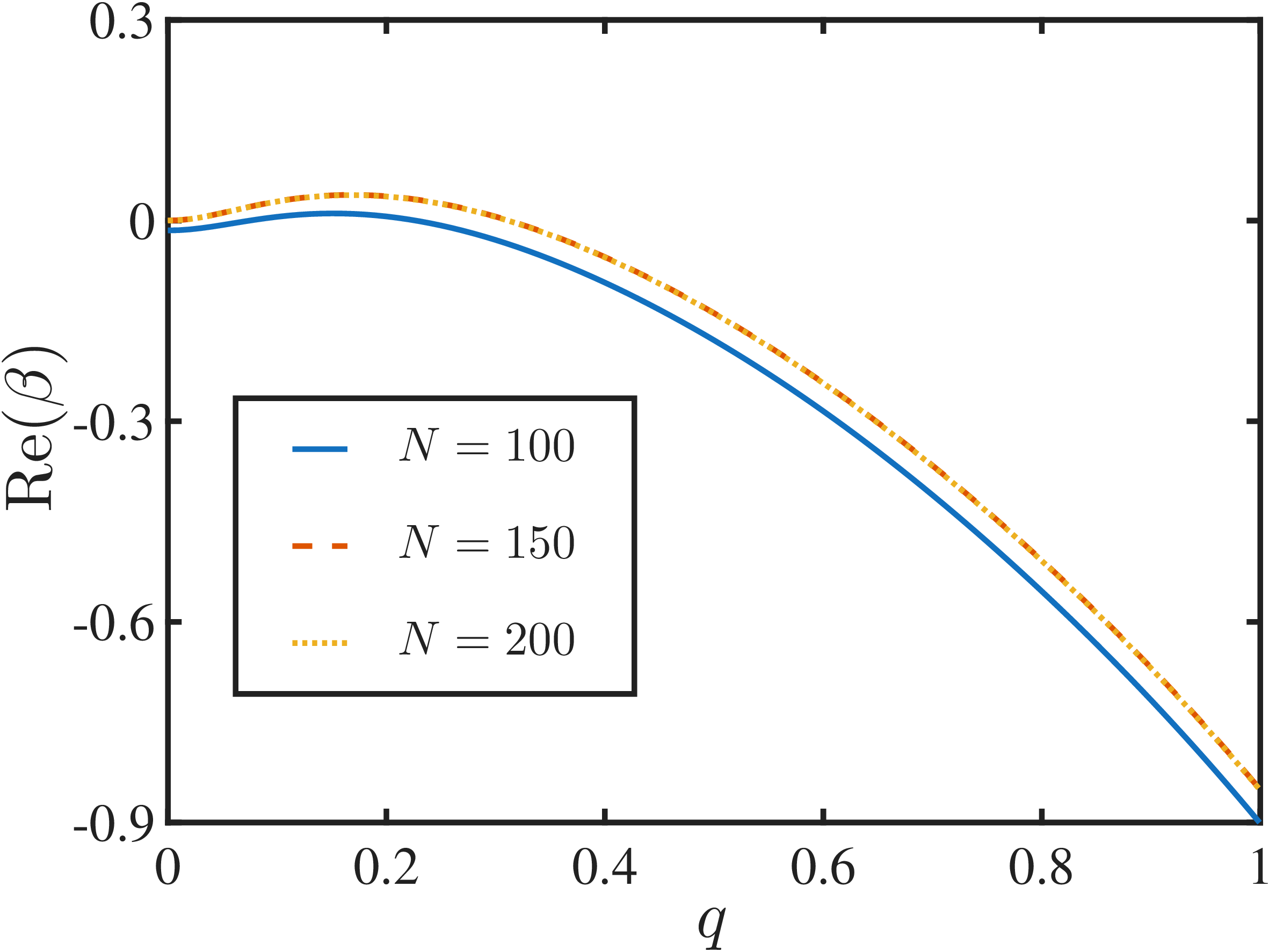}
    \put(-3,70){(b)}
  \end{overpic}
  \caption{Convergence of (a) base flow profile and (b) dispersion curve with respect to the number of collocation points. The corresponding parameters are $\alpha=0.01$, $\phi=\pi/4$, $\theta_e=1$ and $x_f=20$.}
  \label{result:convergence_test_collocation-points}
\end{figure}

\begin{figure}
  \centering
  \begin{overpic}[width=0.45\textwidth]{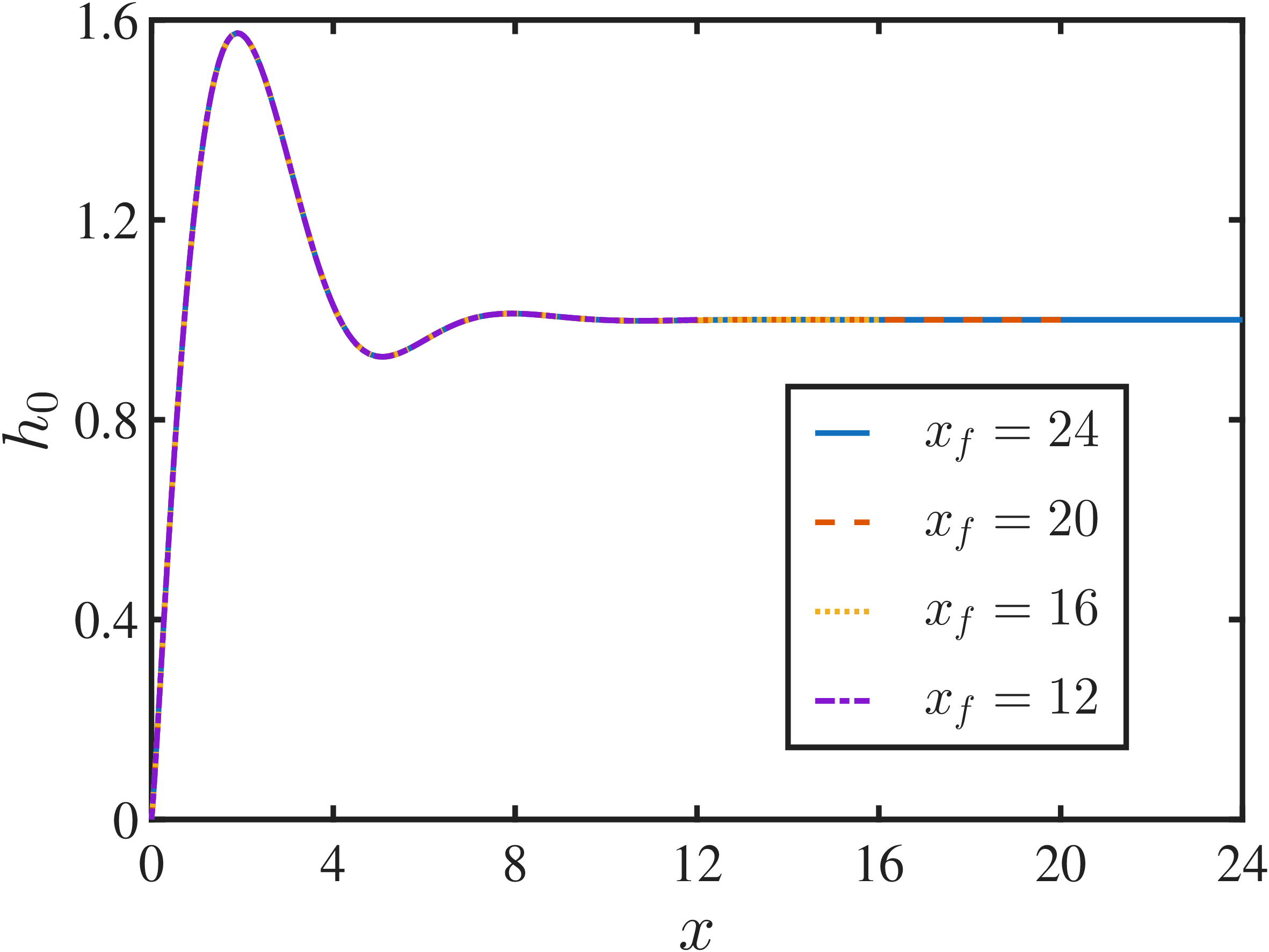}
    \put(-3,70){(a)}
  \end{overpic}
  \quad
  \begin{overpic}[width=0.45\textwidth]{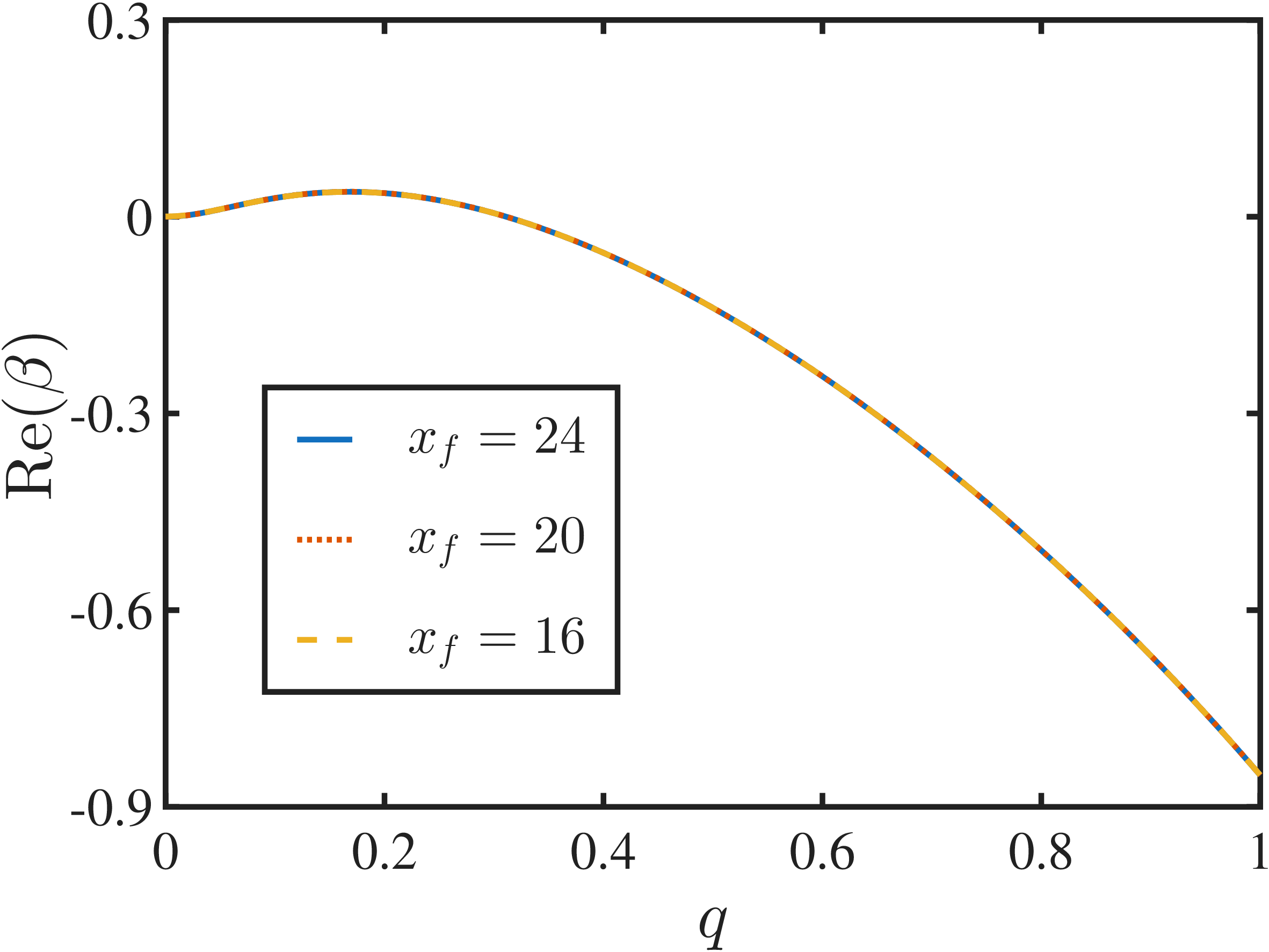}
    \put(-3,70){(b)}
  \end{overpic}
  \caption{Convergence of (a) base flow profile and (b) dispersion curve with respect to the truncated position $x_f$. The corresponding parameters are $\alpha=0.01$, $\phi=\pi/4$, $\theta_e=1$ and $N=300$.}
  \label{result:convergence_test_far-field}
\end{figure}
The semi-infinite domain is truncated at a finite position $x=x_f\gg1$. Since the Chebyshev collocation method is defined on the interval $[-1,1]$, the computational domain must be mapped onto this standard interval. To this end, the physical coordinate ${x}$ is transformed according to  
\begin{equation}
    \tilde{{x}}=\frac{2{x}}{{x}_f}-1,
    \label{eq:mapping}
\end{equation}
where $\tilde{{x}}\in[-1,1]$ denotes the computational coordinate employed in the numerical procedure.

\begin{figure}
    \centerline{\includegraphics[width=0.48\textwidth]{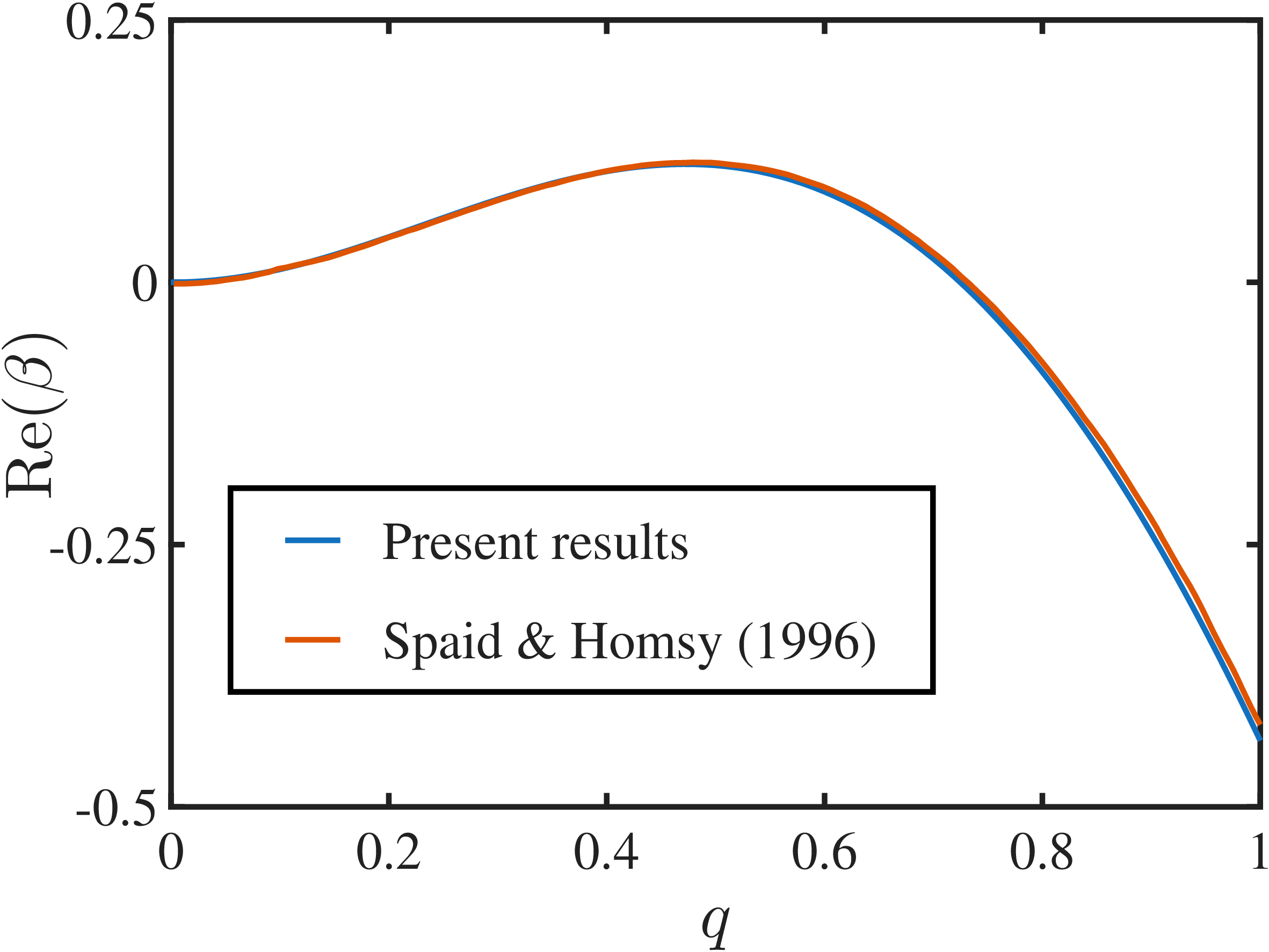}}
    \caption{Comparison of the present dispersion relation with that reported by \citet{spaidStabilityNewtonianViscoelastic1996}. The corresponding parameters are $\alpha=0.1$, $\phi=0$ and $\theta_e=0.05$.}
    \label{fig:comparison_literature}
\end{figure}

Following this transformation, the eigenfunction $G(\tilde{{x}})$ is approximated by a truncated and discretized Chebyshev expansion,
\begin{equation}
    G(\tilde{{x}}_j)=\sum_{n=0}^{N} a_n T_n(\tilde{{x}}_j), \quad j=0,1,\ldots,N,
    \label{eq:chebyshev_expansion}
\end{equation}
where $a_n$ are the expansion coefficients, $N+1$ is the number of collocation points, and $T_n(\tilde{{x}}_j)$ is a $(N+1)\times (N+1)$ matrix determined by the Chebyshev polynomial and collocation points. In the present study, the Chebyshev--Gauss--Lobatto collocation points are employed, which are
\begin{equation}
    \tilde{{x}}_j=\cos\left(\frac{j\pi}{N}\right),
    \quad j=0,1,\ldots,N.
\end{equation}
Similarly, higher-order derivatives of $G(\tilde{{x}})$ at the collocation points can be evaluated through corresponding $(N+1)\times (N+1)$ differentiation matrices. The physical boundary conditions are explicitly imposed by substituting the appropriate rows of the discretization matrices. Notably, imposing boundary conditions in this manner naturally resolves the ill-posedness of Eq.~\eqref{dimensionless-lubri-eq} at the contact line, thereby avoiding the need for an artificial cutoff \citep{wangAxisymmetricEvolutionThin2025} or lifting thickness \citep{DiezGlobalmodelsformovin2000} near the contact line. Through this spectral procedure, the continuous eigenvalue problem Eq.~\eqref{Linearized_perturbation_eq} is ultimately reduced to an algebraic matrix eigenvalue problem.

This numerical implementation was carefully validated through convergence tests. Fig.~\ref{result:convergence_test_collocation-points} presents the convergence of both the base-state profile and the dispersion relation with respect to the number of collocation points. Spectral convergence is achieved for $N=200$. The influence of the computational domain size is examined in Fig.~\ref{result:convergence_test_far-field}, where both quantities are found to be independent of the far-field truncated position once $x_f\ge16$. Throughout the present study, we therefore adopt $N=300$ and $x_f=20$ to ensure fully converged results. Finally, the computed dispersion relation is compared with the results of \citet{spaidStabilityNewtonianViscoelastic1996} in Fig.~\ref{fig:comparison_literature}. The excellent agreement provides further validation of the numerical method.

% \nocite{*}

\bibliography{apssamp}% Produces the bibliography via BibTeX.

\end{document}